\documentclass[twocolumn,showpacs,preprintnumbers,nofootinbib,amsmath,amssymb]{revtex4}
\begin{document}

\title{Multiple M$0$-brane system in an arbitrary eleven dimensional supergravity background}

\author{Igor A. Bandos $^{\dagger\ddagger}$
}
\address{$^{\dagger}$Department of
Theoretical Physics, University of the Basque Country
 UPV/EHU,
P.O. Box 644, 48080 Bilbao, Spain
 \\ $^{\ddagger}$
IKERBASQUE, Basque Foundation for Science, 48011, Bilbao, Spain}

\date{Sept. 14, 2010. V2: Oct. 21, 2010. Misprints corrected, Nov. 24, 2010. Phys. Rev. {\bf D82}, 105030 (2010)}

\def\theequation{\arabic{section}.\arabic{equation}}

\begin{abstract}

The equations of motion of multiple M$0$--brane (multiple M-wave or mM$0$) system in an arbitrary $D=11$ supergravity superspace, which generalize the Matrix model equations for the case of interaction with a generic 11D supergravity background, are obtained in the frame of superembedding approach. We also derive the BPS equations for supersymmetric bosonic solutions of these mM$0$ equations and show that the set of 1/2 BPS  solutions contain a fuzzy sphere modeling M2 brane as well as that the Nahm equation appears as a particular case of the 1/4 BPS equations.

\end{abstract}


\maketitle

\section{Introduction}

More than 15 years ago a concept of M-theory, hypothetical
underlying theory unifying five consistent string models and eleven
dimensional supergravity, appeared \cite{M-theory}. Since that time
many interesting and important results were obtained including the
unexpected applications of the ideas and methods of String/M-theory, first to studying quantum gauge field
theories \cite{AdS/CFT},  including calculating viscosity of
quark-gluon plasma \cite{viscosity}, and, then, to condensed matter
physics, superfluidity and superconductivity \cite{AppAdS/CFT}.

However, the question on fundamental degrees of freedom of M-theory
still remains open. In several occasions it was expressed the opinion
(see e.g. \cite{Howking}) that the present indirect description  of
M-theory is the best what we can have. This is done in terms of its
perturbative and low energy limits, given respectively by the five
consistent 10 dimensional string theories and eleven dimensional
supergravity, by chain of dualities relating these and by the set of
supersymmetric extended objects, super--$p$--branes or, shorter,
$p$--branes (strings for $p=1$, membranes for $p=2$ {\it etc.}; $p=0$ corresponds to particles).
In M-theoretical perspective the most interesting are
ten dimensional fundamental strings (also called F$1$-branes) and Dirichlet $p$-branes
(D$p$-branes) and eleven dimensional M-branes, {\it i.e.} M$p$-branes
with $p=0,2,5$ (see, however, \cite{Tomas+} for recent interest in
lower dimensional branes). These can be described by supersymmetric
solutions of $D$=10 and $D$=11 supergravity \cite{Duff94-Stelle98}, by
the action functionals given by the integrals over their
$(p+1)$--dimensional worldvolumes $W^{p+1}$ (worldvolume actions)
\cite{BST1987,Dpac,B+T=Dpac,blnpst} and also in the frame of
superembedding approach \cite{bpstv,hs96,hs2,Dima99,IB09:M-D}. This,
following the so--called STV (Sorokin--Tkach--Volkov) approach to superparticles and
superstrings \cite{stv, stv+} (see \cite{Dima99} for more references and \cite{vz+} for a related
studies), describes $p$-branes in terms of embedding of its {\it worldvolume
superspace} (${\cal W}^{(p+1|16)}$ for 11D and type II 10D
$p$-branes) into the target superspace  ($\Sigma^{(D|32)}$ for $D$=11 and $D$=10 type II).

As far as an effective description of multiple brane systems is
concerned, it was quickly appreciated that at very low energy the
system of N nearly coincident D$p$--branes (multiple D$p$-brane or
mD$p$ system) is described by maximally supersymmetric $d=(p+1)$
dimensional $U(N)$ Yang--Mills model (SYM model) \cite{Witten:1995im}.
However, in the problem of constructing a complete (more complete) nonlinear supersymmetric action for multiple D$p$-branes,
posed in ninetieth \cite{Tseytlin:DBInA}, only a particular progress
could be witnessed (see \cite{Dima01} for lower dimensional and
lower co-dimensional branes as well as
\cite{Howe+Linstrom+Linus,Howe+Linstrom+Linus=2007} and
\cite{IB09:D0} discussed below).

The so--called `dielectric brane action' proposed by Myers
\cite{Myers:1999ps}, although widely accepted, is purely bosonic, is not Lorentz invariant and
resisted the attempts of its straightforward Lorentz covariant and
supersymmetric generalization all these years \footnote{This does not
look so surprising if we recall that it was derived in
\cite{Myers:1999ps} by a chain of dualities starting from the 10D
non-Abelian DBI  (Dirac--Born--Infeld) action with symmetric trace prescription
\cite{Tseytlin:DBInA} the supersymmetric generalization of which is
also  unknown. }. Recently developed by Howe, Lindstr\"om and Wulff
{\it boundary fermion approach}
\cite{Howe+Linstrom+Linus,Howe+Linstrom+Linus=2007} provides a
supersymmetric and covariant description of Dirichlet branes, but on
the 'pure classical' (or 'minus one quantization') level in the
sense that, to arrive at the description of multiple D-brane system
in terms of the variables similar to the ones in the standard action for a single
D$p$--brane \cite{Dpac,B+T=Dpac} (usually considered as a
classical or quasi-classical action), one has to perform a
quantization of the boundary fermion sector. The complete
quantization of the model \cite{Howe+Linstrom+Linus,Howe+Linstrom+Linus=2007} should produce
not only worldvolume fields of multiple D$p$-brane system but also
bulk supergravity and higher stringy modes. The partial quantization
of only the boundary fermion sector (applying the original
prescription \cite{Marcus+Sagnotti=86} of replacing the boundary
fermion variables by gamma matrices of corresponding internal
symmetry group) allowed Howe, Lindstr\"om and Wulff to reproduce the
purely bosonic Myers action \cite{Myers:1999ps}, but the Lorentz
invariance was lost on this way. See \cite{IB09:D0,mM0=PLB} and
recent \cite{Howe+Linstrom+Linus=2010} for more discussion on
present status of the boundary fermion approach. Finally, a
(possibly approximate but going beyond $U(N)$ SYM) superembedding
description of mD$p$-brane system was proposed and developed for
mD$0$ in \cite{IB09:D0}.

The situation with the description of multiple M-brane systems was
even more complicated: many years it was even unclear what model provides
the description of multiple M$2$ (mM$2$) system at very low energy.
The expected properties of such a model playing for mM$2$ the same
r\^ole as $U(N)$ SYM for mD$p$ were described in
\cite{Schwarz:2004yj} where also problems hampering the way to its
construction were analyzed. In search for solution of these problem
a new ${\cal N}=8$ supersymmetric $d=3$ Chern-Simons plus matter
model based on Filippov 3--algebra \cite{Filippov} instead of Lie
algebra (Bagger--Lambert--Gustavsson or BLG model) was constructed in \cite{BLG}. However,
presently the commonly accepted candidate for the low energy
description of mM$2$ system is a more conventional $SU(N)\times
SU(N)$ invariant ABJM (Aharony--Bergman--Jafferis--Maldacena) model \cite{ABJM}, although this possesses
only ${\cal N}=6$  manifest  $d=3$ supersymmetries. The search for nonlinear
generalization of the BLG model was resulted in a purely bosonic and
Lorentz non-covariant action \cite{Iengo:2008cq} generalizing the
Myers proposition for a multiple bosonic membrane case. The counterpart of
Myers action for purely bosonic limit of multiple M$0$-brane system (also called multiple M-waves, mM$0$ or multiple gravitons) was constructed in \cite{YLozano+=0207}.

Superembedding approach to the {\it multiple M$0$-brane (mM$0$)} system  was proposed
in \cite{mM0=PLB}. In it the relative motion of mM$0$
constituents is described by the maximally supersymmetric $SU(N)$
gauge theory on ${\cal W}^{(1|16)}$ superspace with one bosonic and
sixteen fermionic directions, the embedding of  which into the target
11D superspace ${\Sigma}^{(11|32)}$ is specified by the
superembedding equation (see Eq. (\ref{SembEq=M0}) below).
This latter produces, as its selfconsistency conditions, the dynamical equations of motion for the mM$0$ center of energy degrees of freedom.

This superembedding approach to mM$0$ system provides a covariant generalization of the Matrix model equations with manifest eleven dimensional Lorentz invariance. In
the light of that a single M$0$--brane is dual to the single
D$0$--brane \cite{B+T=Dpac}, the superembedding description of
multiple M$0$ system, constructed and checked on consistency  for
the case of flat target 11D superspace in \cite{mM0=PLB}, provides
the restoration of 11D Lorentz invariance in the (originally ten
dimensional) multiple D$0$ brane system as it was described by the
superembedding approach of \cite{IB09:D0}.

The aim of the present paper is to derive the equations of motion
for multiple M$0$ system in a generic curved supergravity
superspace in the frame of superembedding approach of
\cite{mM0=PLB}.  These equations describe the multiple M$0$ interaction with the 11D
supergravity fluxes and provide the covariant generalization of the Matrix model \cite{Banks:1996vh} for the case of arbitrary supergravity background. The  form of these equations has been
briefly reported in \cite{mM0=PRL}, where it was emphasized the universality of their
structure: when written with indefinite coefficients, these
equations can be reproduced (up to vanishing of a few of the above coefficients) from requirement of $SO(1,1)\otimes SO(9)$ invariance and very few data on the basic fields describing
the relative motion of mM$0$ constituents and on the fluxes which do
interact with them.

Here we give the details on the derivation of the mM0 equations and perform a complete study of the consistency conditions for the superembedding approach. We show that these consistency conditions are obeyed due to the pull--back of the supergravity equations of motion, namely of
the specific projections of the pull--backs of the Rarita--Schwinger and the Einstein equations to ${\cal W}^{(1|16)}$. This  provides a counterpart of the known fact that the D=11 and D=10 supergravity
superspace constraints, and hence the supergravity equations of motion, can be derived from the requirement of $\kappa$--symmetry of the worldvolume action for a single M-brane and D-brane or
fundamental string, respectively.

We also use the superembedding approach to derive the BPS conditions for the supersymmetric pure
bosonic solutions of the equations of motion. In particular, we present
the explicit form of the 1/2 BPS conditions and show that it has a
fuzzy two-sphere solution describing M2-brane as a 1/2 BPS
configuration of the multiple M$0$ system. We also show that the
famous Nahm equation, which also has a fuzzy--two--sphere--related (fuzzy funnel)
solution, appears as a particular case of the 1/4 BPS equation with vanishing four form flux.

The paper is organized as follows. In Section II we present the necessary details on the superspace formulation of D=11 supergravity \cite{CremmerFerrara80,BrinkHowe80} in the notation close to \cite{BdAPV05}. Section III contains a brief review of superembedding approach in its application to a single M$0$ brane, the equations of motion for this supersymmetric object and the description of the intrinsic and extrinsic geometry of the worldline superspace ${\cal W}^{(1|16)}$ embedded in the curved supergravity superspace $\Sigma^{(11|32)}$. Particularly, in section IIIE we present some properties of the relevant projections of the pull--backs of target superspace `fluxes'  (which are 4-form field strength superfield $F_{abcd}=F_{[abcd]}(Z)$,  gravitino field strength superfield $T_{ab}{}^\alpha = T_{[ab]}{}^\alpha (Z)$ and Riemann tensor superfield
$R_{ab}^{cd}= R_{[ab}{}^{[cd]}(Z)$) to the worldvolume superspace, including the relation between them which follows from the Rarita--Schwinger and  Einstein equations of the $D=11$ supergravity. In Section  IV we formulate our proposal for the description of the multiple M$0$ (mM$0$) system by the $SU(N)$ connection  on the $d=1$ ${\cal N}=16$ worldline superspace  ${\cal W}^{(1|16)}$. This superspace, the embedding of which to  $\Sigma^{(11|32)}$ is restricted by the superembedding equation,  describes the center of energy motion of the mM$0$ system, which we discuss in section IVA. In section IVB  we present the constraints on the $d=1$ ${\cal N}=16$ $SU(N)$ connection ($1d$ $16{\cal N}$ SYM supermultiplet) and, in section IVC, derive the dynamical equations of the relative motion of mM$0$ constituents  which follow from these constraints. The BPS equations for supersymmetric bosonic solutions of the mM$0$ equations of motion are presented in Sec. V where we also describe the 1/2 BPS fuzzy sphere solution modeling M$2$ brane by a configuration of mM$0$ system with $N$ constituents, and the appearance of the Nahm equation from the 1/4 BPS equation of mM$0$. Some useful technical details are presented in the Appendices.

\section{Superspace of $D=11$ supergravity}
\setcounter{equation}{0}

M-branes or M-theory super--$p$--branes are extended objects propagating in $D=11$ supergravity superspace $\Sigma^{(11|32)}$. We denote local coordinates of $\Sigma^{(11|32)}$ by ${Z}^{{ {M}}}=
({x}{}^{ {m}}\, ,
{\theta}^{\check{ {\alpha}}})$ ($\check{\alpha}=1,\ldots ,
32$, $m=0,1,\ldots, 9, 10$), with bosonic $x^\mu$ and fermionic ${\theta}^{\check{ {\alpha}}}$,
$$ x^\mu x^\nu= x^\nu x^\mu\; , \quad x^\mu{\theta}^{\check{ {\alpha}}}= {\theta}^{\check{ {\alpha}}}x^\mu\; , \quad {\theta}^{\check{ {\alpha}}}{\theta}^{\check{ {\beta}}}=- {\theta}^{\check{ {\beta}}} {\theta}^{\check{ {\alpha}}}\; .
 $$
 The supergravity is described by the set of supervielbein one forms
\begin{eqnarray} \label{EA=} {E}^{ {A}}:=
dZ^{ {M}}E_{ {M}}{}^{ {A}}(Z)=
 ({E}^{ {a}}, {E}^{ {\alpha}})\; , \qquad
  \end{eqnarray}
including bosonic vectorial form ${E}^{a}$ ($a=0,1,\ldots, 9, 10$) and fermionic spinorial form ${E}^{ {\alpha}}$ (${\alpha}=1,\ldots , 32$), which satisfy the
set of superspace constraints \cite{CremmerFerrara80,BrinkHowe80}. The most important of these constraints determine the bosonic torsion 2--form of  $\Sigma^{(11|32)}$. This reads
\begin{eqnarray}
\label{Ta=11D}  T^{{a}}:= DE^{{a}} =
-i{E}^\alpha \wedge  {E}^\beta \Gamma^{{a}}_{\alpha\beta}\; , \qquad
\end{eqnarray}
where $\Gamma^{{a}}_{\alpha\beta}=\Gamma^{{a}}_{\beta\alpha}$ are 11D Dirac matrices (see Appendix A), $\wedge$ denotes the exterior product of differential forms,
$${E}^b \wedge {E}^a   = - {E}^a \wedge {E}^b\, , \quad {E}^b \wedge {E}^\alpha   = - {E}^\alpha \wedge {E}^b\, , \quad$$ $$ {E}^\beta \wedge {E}^\alpha   = {E}^\alpha \wedge {E}^\beta\; , $$ and $D$ denotes the covariant derivative, $DE^{{a}} =dE^{{a}}-E^b\wedge w_b{}^a$, where
$w{}^{ba}=dZ^Mw_M^{ba}(Z)=-w{}^{ab}$ is the superspace $SO(1,10)$ connection one form (11D spin connection).

After imposing a set of conventional constraints, the study of Bianchi identities  (see \cite{CremmerFerrara80,BrinkHowe80} and, {\it e.g.} \cite{BdAPV05} and refs therein) fixes the form of the fermionic torsion to be
\begin{eqnarray} \label{Tf=}
T^\alpha:= DE^\alpha &=&- E^a \wedge E^\beta
t_{a\beta}{}^{\!\alpha} +
{1 \over 2} E^a \wedge E^b T_{ba}{}^\alpha(Z) \; , \qquad \end{eqnarray}
where
\begin{eqnarray}\label{ta:=}
t_{a\beta}{}^{\!\alpha}&:= {i \over 18}
\left(F_{abcd}\Gamma^{bcd}{}_\beta^{\;\alpha} + {1\over 8}
F^{bcde} \Gamma_{abcde}{}_\beta^{\;\alpha}\right)  \qquad
\end{eqnarray}
is expressed in terms of the fourth rank antisymmetric tensor superfield  $F_{abcd}= F_{[abcd]}(Z)$ ('4-form flux') which obeys
\begin{eqnarray}
\label{D[aFbcde]=0}
D_{[a}F_{bcde]}=0 \; . \qquad
\end{eqnarray}
This indicates that the leading component ($\theta=0$ value) of $F_{abcd}$ can be identified with the field strength of the 3-form gauge field of the 11D supergravity.

Furthermore, the supergravity Bianchi identities  also express the superspace Riemann tensor 2-form
\begin{widetext}
\begin{eqnarray}
\label{RL=} R^{ab}:= (d\omega- \omega\wedge \omega)^{ab} &=&E^\alpha \wedge E^\beta \left( -{1\over3}
F^{abc_1c_2}\Gamma_{c_1c_2} + {i\over 3^. 5!} (\ast
F)^{abc_1\ldots c_5} \Gamma_{c_1\ldots c_5} \right)_{\alpha\beta}
+ \nonumber
\\ &&    +  E^c \wedge E^\alpha \left(
-iT^{ab\beta}\Gamma_{c}{}_{\beta\alpha} + 2i T_c{}^{[a \, \beta}
\Gamma^{b]}{}_{\beta\alpha} \right) + {1 \over 2} E^d \wedge E^c
R_{cd}{}^{ab}(Z) \;
\end{eqnarray}
\end{widetext}
in terms of the same antisymmetric tensor superfield $F_{abcd}(Z)$, the superspace generalization of the gravitino field strength  $T_{ab}{}^\alpha (Z)$ ('fermionic flux' defined in Eq. (\ref{Tf=})) and Riemann tensor superfield $R_{ab}{ }^{cd}=R_{ab}{ }^{cd}(Z)=-R_{ba}{ }^{cd}=-R_{ab}{ }^{dc}$ obeying
\begin{eqnarray}\label{BI:R4}
R_{[ab\; c]}{}^d=0\; . \qquad
\end{eqnarray}
To be convinced that the supervielbein and Lorentz connection obeying the above set of superspace
constraints describe just the supergravity multiplet and no other fields are present, one notices
that the supergravity Bianchi identities also express the fermionic covariant derivatives of the
antisymmetric tensor superfield $F_{abcd}$, of the fermionic flux $T_{ab}{}^\alpha$ and of the
Riemann tensor superfield $R_{cd\,a}{}^{b}(Z)$ through the same set of superfields. In particular,
\begin{eqnarray}
\label{DF4=TG} D_{\alpha}F_{abcd}=- 6\, T_{[ab}{}^\beta \Gamma_{cd]}{}_{\beta\alpha}\; , \qquad
\\
\label{DTabf=RG+Dt+tt} D_{\alpha}T_{ab}{}^{\beta}=-{1\over 4} R_{ab}{}^{cd} \Gamma_{cd}{}_{\alpha}{}^{\beta} - 2(D_{[a}t_{b]} +t_{[a}t_{b]})_{\alpha}{}^{\beta}\; , \qquad \end{eqnarray}
where $t_{a\alpha}{}^\beta$ is expressed through $F_{abcd}(Z)$ by Eq. (\ref{ta:=}).

Further study of Bianchi identities also shows that the superspace constraints (\ref{Ta=11D}) are {\it on-shell}, {\it i.e.} that the supergravity equations of motion appear as their consequences. Those include Einstein equations
\begin{eqnarray} \label{Eq=Einstein}
R_{ab}= - {1\over 3}F_{ac_1c_2c_3}F_{b}{}^{c_1c_2c_3}  +  {1\over 36}\eta_{ab} F_{c_1c_2c_3c_4}F^{c_1c_2c_3c_4}\; , \nonumber \\ \qquad R_{ab}:= R_{ac\, b}{}^c\; , \qquad \eta_{ab}=diag(+,-,\ldots , -)\;  \qquad
\end{eqnarray}
and the Rarita-Schwinger equations $T_{bc}{}^\beta \Gamma^{abc}_{\beta\alpha}=0$. It is convenient to write this latter in the equivalent form of
\begin{eqnarray} \label{Eq=RS}
T_{ab}{}^\beta \Gamma^{b}_{\beta\alpha}=0\; . \qquad
\end{eqnarray}

\section{Superembedding approach to single M$0$-brane and geometry of the worldline superspace ${\cal W}^{(p+1|16)}$}
\setcounter{equation}{0}

\subsection{Superembedding equation}
The standard formulation of M$p$-branes deals with embedding of a purely bosonic worldvolume ${W}^{p+1}$ (worldline ${W}^{1}$ for the case of M$0$-brane) into the {\it target superspace} $\Sigma^{(11|32)}$. The {\it superembedding approach} to M-branes \cite{bpstv,hs2}
describes their dynamics in terms of embedding of  {\it worldvolume superspace} ${\cal W}^{(p+1|16)}$ with $d=p+1$ bosonic and $16$ fermionic directions into  $\Sigma^{(11|32)}$. This embedding can be described in terms of coordinate functions $\hat{Z}^{{ {M}}}(\zeta)=
(\hat{x}{}^{ {m}}(\zeta)\, , \hat{\theta}^{\check{ {\alpha}}}(\zeta))$, which are superfields depending on the local coordinates $\zeta^{{\cal M}}$ of  ${\cal W }^{(p+1|16)}$,
\begin{eqnarray}
\label{WinS} & {\cal W }^{(p+1|16)}\in \Sigma^{(11|32)} : \quad
Z^{ {M}}= \hat{Z}^{ {M}}(\zeta^{{\cal N}}) \;  . \qquad
\end{eqnarray}
For $p=0$ these are  $\zeta^{{\cal N}}=(\tau ,\eta^{\check{q}})$, where  $\eta^{\check{q}}$ are 16 fermionic coordinates  of the {\it worldline superspace} ${\cal W }^{(1|16)}$,
\begin{eqnarray}
\label{WinS} & {\cal W }^{(1|16)}\in \Sigma^{(11|32)} : \quad
Z^{ {M}}=  \hat{Z}^{ {M}}(\tau ,  \eta^{\check{q}}) \;  , \qquad \\ \nonumber & \eta^{\check{q}}\eta^{\check{p}}=- \eta^{\check{p}}\eta^{\check{q}}\, , \qquad \check{q}=1,\ldots ,16 \; ,
\end{eqnarray}
and $\tau$ is its bosonic coordinate generalizing the particle proper time.

To describe a super-$p$-brane,  the coordinate functions $\hat{Z}^{ {M}}(\zeta^{{\cal N}})$ have to satisfy the {\it superembedding equation} which
states that the pull--back $\hat{E}^{{a}}:= d\hat{Z}^{M}(\zeta) E_M^{a}(\hat{Z})$ of the bosonic supervielbein form ${E}^{{a}}:= d{Z}^{M} E_M^{a}(Z)$ to the worldvolume superspace has no fermionic projection. In the case of M$0$--brane this superembedding equation  reads
\begin{eqnarray}
\label{SembEq=M0}
  \hat{E}_{+q}{}^{{a}}:= D_{+q}\hat{Z}^M E_M{}^{{a}}(\hat{Z}) = 0\; ,
\qquad
\end{eqnarray}
where  $D_{+q}$ is a fermionic covariant derivative of ${\cal
W}^{(1|16)}$, $q=1,...,16$ is a spinor index of $SO(9)$ and $+$
denotes the `charge' (weight) with respect to the local $SO(1,1)$
group. In our notation the superscript plus index is equivalent to
the subscript minus, and vice-versa, so that one can equivalently write $D_{+q}$=$D^-_q$.

We denote the supervielbein of $W^{(1|16)}$ by
 \begin{eqnarray}
\label{eA=e0eq} e^{\cal A}= d\zeta^{{\cal M}} e_{{\cal M}}{}^{\cal
A}(\zeta) = (e^{\#}\; , \; e^{+q}) \; ,    \qquad
\end{eqnarray}
and  the only bosonic covariant derivative of ${\cal W}^{(1|16)}$ by $D_{\#}:=D_{++}$
so that  $D=e^{\cal A}D_{\cal A}$ with
\begin{eqnarray}
\label{DA=DbDf}
 D_{\cal A} = (D_{\#}, D_{+q})\; .
\qquad
\end{eqnarray}

\bigskip

\subsection{Moving frame and spinor moving frame variables}

To study the consequences of the superembedding equation, it is
convenient to introduce the auxiliary {\it moving frame superfields}
$u_{{a}}^{=}$, $u_{{a}}^{\#}$, $u_{{b}}^i$ which obey
\begin{eqnarray}\label{u++u++=0}
u_{ {a}}^{=} u^{ {a}\; =}=0\; , \qquad u_{ {a}}^{\# } u^{ {a}\; \#
}=0\; , \qquad    u_{ {a}}^{\; \# } u^{ {a}\; =}= 2\; , \quad
\nonumber \\  \label{uiuj=-}  u_{ {a}}^{=} u^{ {a}\,i}=0\; , \qquad
 u_{{a}}^{\;\#} u^{ {a} i}=0\; , \qquad   u_{ {a}}^{ i} u^{ {a} j}=-\delta^{ij}\; . \quad
\end{eqnarray}
The above constraints imply that the $11\times 11$ matrix
constructed from the columns $u_{{a}}^{=}$, $u_{{a}}^{\#}$ and
$u_{{b}}^i$ ({\it moving frame matrix}) is Lorentz group valued,
\begin{eqnarray}\label{harmUin}
U_b^{(a)}= \left({u_b^{=}+ u_b^{\#}\over 2}, u_b^{i}, {
u_b^{\#}-u_b^{=}\over 2} \right)\; \in \; SO(1,10)\; . \nonumber \\
\end{eqnarray}

To clarify the way these moving frame variables appear in the superembedding approach
let us first notice that the superembedding equation (\ref{SembEq=M0})  can be
written in the form of
\begin{eqnarray}
\label{Eua=e++u--}
  \hat{E}{}^{{a}}:= d\hat{Z}^M(\zeta) E_M{}^{{a}}(\hat{Z}(\zeta)) = {1\over 2}e^{\#}u^{= a}\;
\qquad
\end{eqnarray}
with some 11-vector superfield $u^{=}_{ a}=u^{=}_{ a}(\zeta)$. The
study of consistency conditions shows that this vector must be
lightlike, $u^{=}_{ a}u^{a=}=0$, which allows for its identification
with one of the light--like components of the moving frame
(\ref{harmUin}).

More precisely, the integrability conditions for the superembedding
equation imply that
\begin{eqnarray} \label{Iu--=vGv}  \delta_{qp}
u^{=}_{ a}= v_q^{-\alpha} \Gamma^a_{\alpha\beta} v_p^{-\beta}\;
\qquad
\end{eqnarray}
where the set of $16$ spinorial superfields $v_q^{-\alpha}$ appear
as coefficients for the ${\cal W}^{(1|16)}$ fermionic supervielbein forms in the expressions
for the pull--backs of the target superspace fermionic supervielbein
forms,
\begin{eqnarray} \label{Ef=efv+}
\hat{E}^\alpha :=d\hat{Z}^M(\zeta) E_M{}^\alpha (\hat{Z}(\zeta))=
e^{+q} v_q^{-\alpha} + e^{\#} \chi_{\#}^{-q}  v_q^{+\alpha}\; . \;
\nonumber \\  \qquad
\end{eqnarray}
Then one can show that, as a consequence of (\ref{Iu--=vGv}), the 11--vector superfield $u^{=}_{ a}$ is light-like and finds that it can be completed up the complete  moving frame (\ref{harmUin}).

In a theory with $SO(1,1)\times SO(9)$ symmetry,  the variables $v_q^{-\alpha}$ obeying the constraints (\ref{Iu--=vGv}) parametrize the celestial sphere $S^{9}$ ($9=D-2$ for $D=11$; see
\cite{Ghsds} for $D=4,6,10$ and \cite{GHT93,IB07:M0} for $D=11$ superparticle cases). They form a $32\times 16$ matrix which can be completed till the $32\times 32$ {\it spinor moving frame} matrix
\begin{eqnarray}\label{harmVin}
V_{(\beta)}^{\;\;\; \alpha}= \left(\begin{matrix}  v^{+\alpha}_q
 \cr  v^{-\alpha}_q \end{matrix} \right) \in Spin(1,10)
 \; . \qquad
\end{eqnarray}
(Notice that  $v_{q}^{+\alpha}$ has been already used in  Eq. (\ref{Ef=efv+})).
This  spinor moving frame matrix is related to the moving frame matrix (\ref{harmUin}) by the constraints expressing the Lorentz invariance of the Dirac matrices,
\begin{eqnarray}\label{VGV=GU}
V\Gamma_b V^T =  u_b^{(a)} {\Gamma}_{(a)}\; ,
\qquad \\ \label{VtGV=tGU}
 V^T \tilde{\Gamma}^{(a)}  V = \tilde{\Gamma}^{b} u_b^{(a)} \; , \qquad
\end{eqnarray}
and of the charge conjugation matrix,
\begin{eqnarray}\label{VCV=C}
 VCV^T=C, \qquad V^TC^{-1}V=C^{-1}
 \;  . \qquad
\end{eqnarray}
The relation (\ref{Iu--=vGv}) appears as a $16\times 16$ block in the splitting of the $32\times 32$ matrix of constraint (\ref{VGV=GU}). The constraint (\ref{VCV=C}) allows us to express
the elements of the inverse spinor moving frame matrix
($V_\alpha^{(\beta)}= (v_\alpha{}_q^{-}\; , v_\alpha{}_{q}^{+})\;  \in
\; Spin(1,10)$)  in terms of the original moving frame variables  (\ref{harmVin})
\begin{eqnarray}
\label{V-1=CV}  v_{\alpha}{}^{\mp}_q = \pm i
C_{\alpha\beta}v_{q}^{\mp \beta }\; , \qquad  v^{\pm \alpha}_q = \pm i
C^{\alpha\beta}v_{\beta q}^{\; \pm}\; . \qquad
 \end{eqnarray}
(In our case of $D=11$ with our mostly plus notation the charge conjugation matrix is imaginary, hence the appearance of $i$ in Eqs. (\ref{V-1=CV})).

The moving frame and spinor moving frame variables are also used to construct the $SO(1,1)$ and $SO(9)$ connections on the worldvolume superspace ${\cal W}^{(1|16)}$. The simplest way to define this connection as induced by (super)embedding is to write the $SO(1,10)\times SO(1,1)\times SO(9)$   covariant derivatives (\ref{DA=DbDf}) of the moving frame and spinor moving frame
variables as follows
\begin{eqnarray}
\label{Du--}  Du^{=}_a &=& u^{i}_a \Omega^{=i} \; , \qquad \\ \label{Du++}  Du^{\#
}_a &=& u^{i}_a \Omega^{\# i} \; , \qquad \\ \label{Dui}   Du^{i}_a
 &=& {1\over 2} u^{\# }_a \Omega^{=i}
+ {1\over 2} u^{=}_a\Omega^{\# i} \; ,
\\
\label{Dv-q=}  Dv_q^{-\alpha} &=& - {1\over 2} \Omega^{= i}
\gamma_{qp}^{i}v_p^{+\alpha} \; , \qquad \\
\label{Dv+q=}
 Dv_q^{+\alpha} &=&  - {1\over 2} \Omega^{\# i}
\gamma_{qp}^{i}v_p^{-\alpha} \; . \qquad
\end{eqnarray}
Here $\Omega^{=i}$ and $\Omega^{\# i}$ generalize the ${SO(1,10)\over SO(1,1)\times SO(9)}$ Cartan forms for the case of curved target superspace.

Now the $SO(1,1)$ curvature,  $r=d\omega^{(0)}$, of the worldline superspace ${\cal W}^{(1|16)}$ and  the  $SO(9)$ curvatures of the normal bundle over it, ${\cal G}^{ij}$, can be defined through the Ricci identities specified for the moving frame variables,
\begin{eqnarray}
\label{DDu++=} DDu_{ {a}}^{\# } = \; \; 2 d\omega^{(0)}u_{ {a}}^{\#
} + \hat{R}_{ {a}}{}^{ {b}}u_{ {b}}^{\# }\; , \qquad  \\
\label{DDu--=} DDu_{ {a}}^{=} = -2 d\omega^{(0)}u_{ {a}}^{=} +
\hat{R}_{ {a}}{}^{ {b}}u_{ {b}}^{=} \; , \qquad  \\ \label{DDui=}
DDu_{ {a}}{}^{i} = u_{ {a}}^{j} {\cal G}^{ji} + \hat{R}_{ {a}}{}^{
{b}}u_{ {b}}{}^{i} \; . \qquad
\end{eqnarray}
Here $ \hat{R}_{ {a}}{}^{{b}}$ is the pull--back of the target superspace Riemann curvature two form (\ref{RL=}) to ${\cal W}^{(1|16)}$. Contracting Eq. (\ref{DDu++=}) with $u^{=a}$ and Eq.  (\ref{DDui=})  with
$u^{ja}$, and denoting the moving frame projections of the Riemann curvature pull--back $\hat{R}_{ {a}}{}^{{b}}$ by
\begin{eqnarray}\label{M0:hR--i=}
 \hat{R}{}^{=\; \# }:=  \hat{R}^{ {a} {b}}u_{ {a}}^{=} u_{ {b}}^{\# } \; , \qquad
 \hat{R}{}^{ij}:=  \hat{R}^{ {a} {b}}u_{ {a}}^{i} u_{ {b}}{}^j  \;
,   \qquad \nonumber \\
 \hat{R}{}^{=j}:=  \hat{R}^{ {a} {b}}u_{ {a}}^{=} u_{ {b}}{}^j  \; ,  \qquad
 \hat{R}{}^{\# j}:=  \hat{R}^{ {a} {b}}u_{ {a}}^{\#} u_{ {b}}{}^j  \; ,  \qquad
\end{eqnarray}
one finds the following generalization of the Gauss and Ricci equations of the Classical Surface Theory
(see \cite{bpstv} for references)
\begin{eqnarray}\label{M0:Gauss}
 d\omega^{(0)} &= & {1\over 4 } \hat{R}{}^{=\; \# }+  {1\over 4 } \Omega^{=\, i} \wedge
 \Omega^{\# \, i}\; , \qquad  \\
\label{M0:Ricci} {\cal G}^{ij} & =& \hat{R}{}^{ij}-   \Omega^{=\,
[i} \wedge \Omega^{\# \, j]}\;
.  \qquad
\end{eqnarray}
One can also use (\ref{DDu++=}) and (\ref{DDu--=}) to obtain, as integrability conditions of Eqs. (\ref{Du--}) and (\ref{Du++}), the following generalization of the Peterson--Codazzi equations
\begin{eqnarray}\label{M0:DOm--=} D\Omega^{= i} =  \hat{R}{}^{= i}\; , \qquad
D\Omega^{\# i} =  \hat{R}{}^{\# i}\;  .  \qquad
\end{eqnarray}

More details on moving frame variables and their role in superembedding approach
can be found in Appendix B as well as in \cite{mM0=PLB,IB09:M-D} in the case of M$0$-brane  and in  \cite{bpstv,Dima99,IB09:M-D} (and in refs therein) in the general case.

\subsection{Equations of motion of a single M$0$-brane from superembedding approach}

The superembedding equation (\ref{SembEq=M0})  is {\it on--shell} in
the sense that it contains the M$0$-brane equations of motion among
its consequences. We refer to  \cite{mM0=PLB} for the details on the derivation
of these equations and just present the result. The fermionic equations of motion state the vanishing of the bosonic component of the pull--back of fermionic supervielbein of ${\cal W}^{(1|16)}$,
\begin{eqnarray}\label{M0:DiracEq}
\chi_{\# p}{}^-&:=& \hat{{E}}_{\# }{}^{ {\alpha}}v_{ {\alpha}p}{}^-
=0 \;  , \qquad  \end{eqnarray} so that on the mass shell Eq. (\ref{Ef=efv+}) simplifies to
\begin{eqnarray} \label{Ef=efv}
\hat{E}^\alpha :=d\hat{Z}^M(\zeta) E_M{}^\alpha (\hat{Z}(\zeta))=
e^{+q} v_q^{-\alpha} \; .   \qquad
\end{eqnarray}
The bosonic equation of motion for M$0$--brane reads
\begin{eqnarray}\label{M0:bEq}
\Omega_{\# }^{\; = i}\, &:=& -D_{\# } u^{=  {a}}\, u_{ {a}}{}^i = -
D_{\# } \hat{E}_{\# }{}^{  {a}}\, u_{ {a}}{}^i=0 \; . \qquad
\end{eqnarray}
Together with (\ref{Ef=efv}), Eq. (\ref{M0:bEq}) implies the
differential form equation
\begin{eqnarray}\label{M0:1-formEq}
\Omega^{\; = i}:=  - Du^{a=} \; u_a{}^{i}= 0 \;  \qquad
\end{eqnarray}
stating vanishing of the one-form in the {\it r.h.s}'s of Eqs. (\ref{Du--}) and (\ref{Dv-q=}).
Hence the dynamical equations of M$0$-brane can be
formulated as the condition that the lightlike  moving frame vector $u_a^{=}$ and
its square root (in the sense of Eq. (\ref{Iu--=vGv})), the set of 16 constrained spinorial superfields  $v_q^{-\alpha}$,  are covariantly constants,
\begin{eqnarray} \label{M0:Eq=Du--=0}
Du_a^{=}=0 \; , \qquad Dv_a^{-\alpha}=0 \; . \qquad
\end{eqnarray}

\subsection{Geometry of the the worldline superspace ${\cal W}^{(1|16)}$ and of the $SO(9)$ bundle over it}

As far as the worldline superspace ${\cal W}^{(1|16)}$  whose embedding into the target eleven dimensional superspace $\Sigma^{(11|32)}$  will be used to describe the motion of the multiple M$0$ system, we will need some details on the  geometry of ${\cal W}^{(1|16)}$  and of the normal bundle over it.

Taking into account  Eqs. (\ref{M0:DiracEq}) and  (\ref{M0:bEq}), one finds that, similarly to the case of flat superspace, the bosonic torsion two form of  ${\cal W}^{(1|16)}$  is given by
\begin{eqnarray}
\label{De++=efef} De^{\# }=-2ie^{+ q} \wedge  e^{+q}\; , \qquad
\end{eqnarray}
and that the curvature of the $SO(1,1)$ connection on $W^{(1|16)}$ vanishes
\begin{eqnarray}
\label{dOm0=0}
d\omega^{(0)}=0\;  \qquad
\end{eqnarray}
(see Gauss equation (\ref{M0:Gauss})). Nevertheless, the geometry induced on $W^{(1|16)}$  by its embedding to $\Sigma^{(11|32)}$ is not trivial because the fermionic torsion two form is nonzero,
\begin{eqnarray} \label{M0:De+q=}
De^{+q}=- {1 \over 72} e^{\# } \wedge e^{+p} \hat{F}_{\# ijk}\gamma^{ijk}_{pq}  \; . \qquad
\end{eqnarray}
Here $i,j,k=1,\ldots , 9$, $\; \gamma^{ijk}= \gamma^{[i}\gamma^{j}\gamma^{k]}$ is the antisymmetric product of the
nine-dimensional Dirac matrices,  $\gamma^i_{qp}= \gamma^i_{pq}$, obeying
\begin{eqnarray} \label{d9gamma}\gamma^i\gamma^j+ \gamma^j\gamma^i =\delta^{ij}I_{\!_{16\times 16}}\; , \quad i,j=1,\ldots , 9
\end{eqnarray} (some useful properties of these can be found in Appendix A). Eq. (\ref{M0:De+q=}) expresses the fermionic torsion in terms of the projection
\begin{eqnarray}\label{M0:Fluxes}
 \hat{F}_{\# ijk}:= F^{{a}{b}{c}{d}} (\hat{Z}) u_{{a}}
{}^{=}u_{{b}}{}^{i}u_{{c}}{}^{j}u_{{d}}{}^{k} \;  \qquad
\end{eqnarray}
of the pull--back to  ${\cal W}^{(1|16)}$ of the 4-form flux (4-form field strength superfield) $F^{{a}{b}{c}{d}}({Z})$ of the eleven dimensional supergravity. This flux projection enters as well in the expression for the $SO(9)$ curvature of normal bundle over ${\cal W}^{(1|16)}$ determined by Ricci equation (\ref{M0:Ricci}),
\begin{widetext}
\begin{eqnarray} \label{M0:Gij=onshell}
{\cal G}^{ij}= \hat{R}^{ij}= e^{+q}\wedge e^{+p} \left( {2i\over 3}\hat{F}_{\#\, ijk}\gamma^{k}_{qp}+ {i\over 18}\hat{F}_{\#\, klm}\gamma^{ijklm}_{qp} \right) - i  e^{\# }\wedge e^{+q} \gamma_{qp\; [i}\hat{T}_{\#\, j]+p} \; . \qquad
\end{eqnarray}
\end{widetext}
The last term in (\ref{M0:Gij=onshell}) contains the projection
\begin{eqnarray}
\label{M0:fFluxe} \hat{T}_{\#\, i\, +q} :=T_{{a}{b}}{}^{{\beta}}(\hat{Z}) v_{{\beta}q}^{\; -}\,
u_{{a}}^{=}u_{{b}}^i\;  \qquad
\end{eqnarray}
of the pull--back to $W^{(1|16)}$ of the `fermionic flux'   $T_{{a}{b}}{}^{{\beta}}({Z})$ (superfield generalization of the gravitino field strength, see Eq. (\ref{Tf=})).

Here and below, to make equations lighter, we identify upper and
lower case SO(9) vector indices; although our 11D metric is 'mostly
minus', $\eta^{ab}=diag(+,-,...,-)$, this should  not
produce a confusion as far as we never use contractions of 'internal' indices with
$\eta_{ij}=-\delta^{ij}$. We also conventionally replace $^=$ superscript by $_\#$ subscript in the notation for the contractions of the tensors with $u_a^=$.

Notice that, with the M$0$ equations of motion written in the form of Eq. (\ref{M0:1-formEq}), $\Omega^{= i}=0$, the Peterson--Codazzi equation (\ref{M0:DOm--=}) results in $$\hat{R}{}^{ =i}:=  \hat{R}^{ {a} {b}}u_{ {a}}^{=} u_{ {b}}{}^i=0\; . $$ Calculating the pull--back of the Riemann curvature two form (\ref{RL=}) to ${\cal W}^{(1|16)}$, one sees  that this relation is satisfied identically.

While $\Omega^{= i}=0$ encodes the M$0$ equations of motion, the second set of $SO(1,1)\otimes SO(9)\otimes SO(1,10)$ covariant one forms $\Omega^{\# i}$ determining the $SO(1,1)\otimes SO(9)\otimes SO(1,10)$ covariant derivatives of $u^{\#}_a$ and $v^{+\alpha}_q$ in Eqs. (\ref{Du++}) and (\ref{Dv+q=}), remains unspecified by the superembedding equations. This reflects the $K_{9}$ gauge symmetry of the massless superparticle dynamics; in our superembedding approach  $K_{9}$ appears as a gauge symmetry leaving invariant $u^{=}_a$ and $v^{-\alpha}_q$ while acting on the remaining moving frame superfields by  \begin{eqnarray} \label{M0:K9sym} \delta u^{\#}_a = 2k^{\# i} u^i_a\, , \quad \delta u^i_a=  2k^{\# i} u^{=}_a\, , \quad  \delta v^{+\alpha}_q= k^{\# i} \gamma^i_{qp}v^{-\alpha}_q\; . \end{eqnarray}
With respect to $K_9$ the one form $\Omega^{\# i}$ is not covariant but transforms as a connection; actually it can be considered as a part of the connection of a normal bundle over ${\cal W}^{(1|16)}$. The structure group of this normal bundle  is nonstandard,  $SO(9)\subset\!\!\!\!\!\!\times K_9$ (rather than, say $SO(10)$) because the bosonic body of ${\cal W}^{(1|16)}$ is a light-like line in spacetime. However, for our purposes here it is sufficient to account for the $SO(9)$ part of the curvature of this normal bundle and to keep manifest only the $SO(1,1)\otimes SO(9)\otimes SO(1,10)$ gauge symmetry, thus leaving  $K_9$ symmetry hidden.

\subsection{Pull--back of the fluxes  to ${\cal W}^{(1|16)}$ and supergravity equations of motion.}

Thus the characteristics of the geometry of ${\cal W}^{(1|16)}$, induced by its embedding to $\Sigma^{(11|32)}$, and of the normal bundle over it, involve only definite projections (\ref{M0:Fluxes}) and (\ref{M0:fFluxe}) of the pull--backs to ${\cal W}^{(1|16)}$ of the covariant bosonic and fermionic superfields (`fluxes') of the eleven dimensional supergravity. Then, if some model is defined on ${\cal W}^{(1|16)}$, its interaction with background supergravity will be described by this projections of the fluxes and by their derivatives. This poses the problem of calculating the worldline covariant derivatives of superfields (\ref{M0:Fluxes}) and (\ref{M0:fFluxe}) which might seem to be quite involved. Fortunately, the properties of ${\cal W}^{(1|16)}$ simplify these calculations essentially.

Firstly, let us observe that Eqs. (\ref{M0:Fluxes}) and (\ref{M0:fFluxe}) involve only $u^{=}_a$, $v_{\alpha q}^-$ and $u^i_a$ moving frame superfields. Then, as it was mentioned above, the  equations of motion for single M$0$--brane (which follow from superembedding equation)  can be expressed by the statement that  $Dv_{\alpha q}^{\; -} =0$ and $Du_a^{=}=0$, Eq. (\ref{M0:Eq=Du--=0}). Furthermore, due to the same equations which can be written in the form of Eq. (\ref{M0:1-formEq}), the derivative of the $u^i_a$ superfield reads
 $Du^i_a ={1\over 2} \Omega^{\# i} u_a^=$ (see (\ref{Dui})). It is important that $Du^i_a\propto  u_a^=$
and, hence, do not contribute in the derivative of an expression constructed from an antisymmetric tensor of $SO(1,10)$ contracting one of its indices with $u^i_a$ and another with $u_a^=$. The projections (\ref{M0:Fluxes}) and (\ref{M0:fFluxe}) of the bosonic and fermionic fluxes  are just of this type so that the calculation of their worldline superspace fermionic covariant derivatives is basically reduced to the algebraic operation with the expressions for the background superspace spinorial derivatives of the corresponding superfields, Eqs. (\ref{DF4=TG}) and (\ref{DTabf=RG+Dt+tt}).

After some algebra using the properties of moving frame and spinor moving frame variables (Eqs. (\ref{vv=uG-all}) and (\ref{V-1=CV-APP}) in  Appendix B), we find that Eq. (\ref{DF4=TG}) implies
\begin{eqnarray} \label{DFijk=}
D_{+q}\hat{F}_{\# ijk}=  3i \gamma_{[ij|\; qp} \hat{T}_{\#\, |k]\; p}
   \;     \qquad
\end{eqnarray}
and Eq. (\ref{DTabf=RG+Dt+tt}) results in
\begin{widetext}
\begin{eqnarray} \label{DpT--iq=}
D_{+p} \hat{T}_{\# \, i\, + q}= {1\over 2} \hat{R}_{\# ij
\#} \gamma^j_{pq} + {1\over 3} D_{\#} \hat{F}_{\# ijk} \left(\delta^{i[j}\gamma^{kl]}_{pq} + {1\over 6}
\gamma^{ijkl}_{pq} \right)+  \hat{F}_{\# j_1j_2j_3}\hat{F}_{\# k_1k_2k_3} {\Sigma}{}^{i\, , \,  j_1j_2j_3\, ,\, k_1k_2k_3}_{pq}
  \;      \qquad
\end{eqnarray}
Here
\begin{eqnarray} \label{M0:RFlux}
& \hat{R}_{\# ij
\#}:= R_{dc\; ba}(\hat{Z})u^{d=}u^{ci} u^{bj}u^{a=} \;    \qquad
\end{eqnarray}
is the specific projection of Riemann tensor and the  explicit form of the last term reads
\begin{eqnarray}\label{FFS=}
  && \hat{F}_{\# j_1j_2j_3}\hat{F}_{\# k_1k_2k_3} {\Sigma}{}^{i\, , \,  j_1j_2j_3\, ,\, k_1k_2k_3}_{pq}=-{1\over 12} \gamma^j_{pq} \left( \hat{F}_{\# i k_1k_2} \hat{F}_{\# j k_1k_2}+{1\over 9}\delta^{ij} (\hat{F}_{\# k_1k_2k_3})^2\right)+ \qquad \nonumber \\ && +{1\over 9} \gamma^{j_1j_2j_3}_{pq}  \hat{F}_{\# i j_1k} \hat{F}_{\# k j_2j_3}  + {1\over 72} \gamma^{ k_1 k_2 k_3 k_4 k_5}_{pq} \left( \hat{F}_{\# i k_1k_2} \hat{F}_{\# k_3k_4k_5} +\delta^{i}_{[k_1} \hat{F}_{\# k_2k_3|j}\hat{F}_{\# j|k_4k_5]}\right)
  \; ,     \;
\end{eqnarray}
\end{widetext}

Notice that the projection (\ref{M0:RFlux}) of the Riemann tensor is symmetric as far as
\begin{eqnarray}\label{hatR[]=0}
\hat{R}_{\#[i\; j]\#}= {3\over 2}\hat{R}_{[\# i\; j]\#}=0  \;    \qquad
\end{eqnarray}
due to Eq. (\ref{BI:R4}),  $R_{[abc]d}=0$. Furthermore, its trace (on SO(9) vector indices)  is expressed through the product of the projections (\ref{M0:Fluxes}) of the 4-form fluxes by
\begin{eqnarray}\label{hatEiEq=singlet}
\hat{R}_{\# j\# j} + {1\over 3} (\hat{F}_{\# ijk})^2= 0 \; , \qquad
\end{eqnarray}
which is the $u^{=}_au^{= b}$ projection of the pull--back of
the supergravity Einstein equation (\ref{Eq=Einstein}) to ${\cal W}^{(1|16)}$.

The contraction of the pull--back to ${\cal W}^{(1|16)}$ of the supergravity Rarita-Schwinger equations (\ref{Eq=RS}) with $u^{--a}v^{-\alpha}_q$ gives
\begin{eqnarray}\label{hatRS=}
\gamma^{i}_{qp}\hat{T}_{\# i \, +p}=0
\; . \qquad \end{eqnarray}
It should not be too surprising that the selfconsistency condition for this equation is satisfied identically when the consequence  (\ref{hatEiEq=singlet}) of the supergravity Einstein equation (\ref{Eq=Einstein}) is taken into account,
$\gamma^{i}_{qs}D_{+p}\hat{T}_{\# i \, +s}= - {1\over 2} \delta_{qp} \left( \hat{R}_{\# j\# j} + {1\over 3} (\hat{F}_{\# ijk})^2\right)= 0$.

Now we have all necessary details on the geometry of the worldline superspace ${\cal W}^{(1|16)}$ induced by its superembedding in ${\Sigma}^{(11|32)}$ and are ready to study the supersymmetric gauge theory on this superspace which we use to describe the relative notion of the constituents of the multiple M$0$--brane  (mM$0$) system.

\section{multiple M$0$ description by $SU(N)$ SYM on ${\cal W}^{(1|16)}$ superspace}
\setcounter{equation}{0}

The superembedding approach to multiple M$0$-brane system implies, in particular, a superfield description of  the relative motion of M$0$ constituents. Our proposition is  to describe the relative motion of M$0$ constituents by the maximally supersymmetric $SU(N)$ YM gauge theory on ${\cal W}^{(1|16)}$ whose embedding into the target 11D superspace is specified by the superembedding equation (\ref{SembEq=M0}) \cite{mM0=PLB}. To motivate such a choice, we firstly notice that, as far as M$0$-brane is dual to type IIA D$0$-brane \cite{B+T=Dpac}, it is natural to expect that multiple M$0$ system is dual to the multiple D$0$-brane one. Then, the worldline superspace $SU(N)$ SYM description of the relative motion in multiple M$0$-system is suggested by the superembedding description of the multiple D$0$'s \cite{IB09:D0}. The suggestion to describe this by a $d=1$ ${\cal N}=16$ $SU(N)$  SYM model on the ${\cal W}^{(1|16)}$ superspace comes from the fact that at very low energy the gauge fixed description of the dynamics of the  multiple D$p$-brane system in flat target type II superspace can be provided by maximally supersymmetric $(p+1)$--dimensional $U(N)\,$ ($\,=SU(N)\otimes U(1)$) SYM model, {\it i.e.} by dimensional reduction of the corresponding $D=10$ SYM model with $U(N)$ gauge symmetry
\cite{Witten:1995im}.

Now we have to specify the embedding of the 'center of mass' (better to say, 'center of energy')  superspace ${\cal W}^{(1|16)}$ of the multiple M$0$ system into the target superspace $\Sigma^{(11|32)}$ of eleven dimensional supergravity. The natural proposition is to require this to be defined by the superembedding equation (\ref{SembEq=M0}). The arguments in favor of such a choice include the universality of the superembedding equation and the difficulty one meets in an attempt to generalize it. Now we can also refer on that the approach based on the use of the center energy superspace ${\cal W}^{(1|16)}$ obeying the superembedding equation was checked on consistency for multiple M$0$ system in flat $D=11$ superspace \cite{mM0=PLB}. However, it was clear from the very beginning that this  superembedding approach is able  to provide a covariant generalization of the matrix model equation valid in any curved 11D supergravity background. In this Section we derive  the explicit form of such equations describing the multiple M$0$ interaction with the 11D supergravity fluxes.

\subsection{mM$0$ center of energy motion from superembedding of ${\cal W}^{(1|16)}$ into ${\Sigma}^{(11|32)}$}

Thus the center of energy superspace of the mM$0$ system is chosen to be  ${\cal W}^{(1|16)}$, the counterpart of the worldline superspace of single M$0$, the embedding of which into the target superspace, an arbitrary 11D supergravity superspace  ${\Sigma}^{(11|32)}$, is restricted by the superembedding equation (\ref{SembEq=M0}). As far as the superembedding equation specifies completely the geometry of the worldline superspace, all the knowledge on the torsion forms and curvature of
${\cal W}^{(1|16)}$ and normal bundle over it, Eqs. (\ref{De++=efef})--(\ref{M0:De+q=}) and (\ref{M0:Gij=onshell}), on its extrinsic geometry, Eq. (\ref{M0:1-formEq}), as well as on the pull--backs of fluxes to ${\cal W}^{(1|16)}$, Eqs. (\ref{M0:Fluxes}) and (\ref{M0:fFluxe})--(\ref{hatRS=}), are true for this center of energy superspace. In particular, the pull--backs of the target space supervielbein to  ${\cal W}^{(1|16)}$ obey Eqs.  (\ref{M0:DiracEq}) and (\ref{M0:bEq}),
which encodes the dynamical equations of motion for single M$0$-brane (equivalent to Eqs. (\ref{M0:Eq=Du--=0})),
\begin{eqnarray}\label{Ea=e+u-2}
& \hat{E}^a =
e^{\#} u^{a=}/2\; , \quad Du^{a=}=0\; , \qquad u^{a=}u_a^{=}=0\; , \quad
 \\ \label{Ef=efv-2}
 & \hat{E}^\alpha =
e^{+q} v_q^{-\alpha}  \; ,   \quad  Dv_q^{-\alpha}=0\; ,  \quad   v_q^{-} \Gamma^a v_p^{-}= \delta_{qp}
u^{=}_{ a}\; . \quad
\end{eqnarray}

The fact that equations of motion for the center of energy of the multiple $p$--brane system have the form of equations for single brane looks natural, in particular when we are speaking about system of particles. However, one have to stress that for the mM$0$ system, as far as single M$0$ brane is a {\it massless} 11D superparticle, the statement that the dynamics of the center of energy is governed by a single M$0$ equations implies that the mM$0$ center of energy moves on a light--like geodesic in the bosonic body of ${\Sigma}^{(11|32)}$.
This fact, expressed by the third equation in (\ref{Ea=e+u-2}) (or, equivalently, by $\hat{E}{}^a_{\#}\hat{E}_{\# a}=0$), should not look surprising if we keep in mind the image of, for instance, a beam of light, which moves as a whole in a light-like direction despite, say, gravitational interaction among photons.

One may also find this property natural for a generalization of Matrix model. Indeed, making a dimensional reduction of our mM$0$ system to 10D, on the way similar to passing from single M$0$ to single D$0$ in \cite{B+T=Dpac} by generalized dimensional reduction, we will find a time-like motion of the center of mass of the 10D system, which would be the mD$0$ system in a type IIA supergravity background. Actually such a system, but in a simpler background, was the final `destination' of the  DLCQ (discrete light-cone quantization) approach in  \cite{Susskind:1997cw,Seiberg:1997ad}
\footnote{
Certainly we appreciate differences between our approach and DLQG reasonings of
\cite{Susskind:1997cw,Seiberg:1997ad} which discusses the M-theory compactification on a light-like circle,
considering this as limit of spacial circle of radius $R_s$ and restricting to the sector of
fixed momentum along the circle $p=N/R_s$ which is argued to produce a theory of N D$0$ branes.
The most evident difference is that in DLCQ the number of D$0$-branes is defined by the integer number characterizing the fixed value of the momentum in the compact direction, $p=N/R_s$, while in our construction  the number of D$0$'s in mD$0$ system is defined by the
the number of mM$0$ constituents in the prototype 11D system and the fixed momentum in compact (space-like) dimension corresponds to the mass of the 10D mD$0$ system.}

\subsection{Basic superfields describing relative motion of mM$0$ constituents and basic constraints for them}

Thus our center of energy superspace ${\cal W}^{(1|16)}$ is defined by the superembedding equation (\ref{SembEq=M0}) imposed on the coordinate functions $\hat{Z}^{{ {M}}}(\zeta)= (\hat{x}{}^{ {m}}(\zeta)\, ,
\hat{\theta}^{\check{ {\alpha}}}(\zeta))$. This results in dynamical equations which formally coincide with the equations of motion of a single M$0$--brane, which implies, in particular, that the center of energy motion is light-like. Our proposition is to describe the relative motion of the mM$0$ constituents by $d=1$, ${\cal N}=16$ $SU(N)$ SYM model on this superspace. This is formulated in terms of 1-form gauge potential $A=e^{\# }A_{\# }+ e^{+q} A_{+q}$ the field strength of which,
\begin{eqnarray}\label{M0:G2:=}
 G_2&=& dA - A\wedge A= \qquad \nonumber \\ &=& 1/2\, e^{+q}\wedge e^{+p} G_{+p\, +q} +  e^{\#}\wedge e^{+q} G_{+q\, \#} \; , \qquad
\end{eqnarray}
should be restricted by the set of constraints. The natural choice for the these constraints is
\begin{eqnarray}\label{M0:G=sX}
G_{+q\, +p}=  i \gamma^i_{qp} {\bf X}{}^i \; ,  \qquad
\end{eqnarray}
where $\gamma^i_{qp}= \gamma^i_{pq}$ are nine-dimensional Dirac
matrices (\ref{d9gamma}) and
${\bf X}{}^i  =-({\bf X}{}^i)^\dagger$ is a nanoplet of N$\times$N anti-hermitian matrix
superfields. The leading component of this, ${\bf X}{}^i \vert_{\eta^q=0}$, provides a natural candidate for the field describing the relative motion of the M0 constituents. As it was stressed in \cite{mM0=PRL}, it is important that this superfield has the $SO(1,1)$ wait 2, the fact which we find convenient to present in the form ${\bf X}{}^i ={\bf X}{}_{\#}^i:={\bf X}{}_{++}^i$ (and which can be seen from Eq. (\ref{M0:G=sX})).

Let us notice that the essential constraint in (\ref{M0:G=sX}) is
$G_{+q\, +p}\gamma^{ijkl}_{pq}=0$, while the vanishing of the $SO(9)$ singlet part of this,
$G_{+q\, +q}=0$, is the conventional constraint which determines $A_{\#}$ in terms of $A_{+q}$ and its derivatives. One can also think about a more complicated set of constraints
$G_{+q\, +p}=  i \gamma^i_{qp} {\bf X}{}^i+ i \gamma^{ijkl}_{qp} {\bf Y}{}^{ijkl}$
where ${\bf Y}{}^{ijkl}$ is constructed from ${\bf X}{}^i$ superfields and their covariant fermionic (and bosonic) derivatives. For the case of flat target superspace this corresponds to a deformation of the $D=10$ SYM model reduced to $d=1$; such deformations do exist and were the subject of studies
in \cite{Cederwall:2001td} and more recent \cite{Movshev:2009ba}. Although the existence of their counterparts corresponding to the curved target superspace seems to be a reasonable conjecture, to our best knowledge no special study of that has been carried for today. Furthermore, even if this conjecture were proved, so that it were natural to expect the appearance of such type deformations in a multiple brane models, the study of such models would promise to be very complicated (up to not being practical, at least without use of a computer programs like the one applied in \cite{Cederwall:2001td}). So in this paper we restrict ourself by considering the model with the simplest constraints (\ref{M0:G=sX}); if the above mentioned deformation were found, our results based on constraint (\ref{M0:G=sX}) would provide at least a reasonable (handible) approximation to such a more complete but much more complicated description.

Studying Bianchi identities $DG_2=0$ one finds that the  selfconsistency
of the constraints (\ref{M0:G=sX}) requires the
matrix superfield ${\bf X}{}^i $ to obey the {\it superembedding--like
equation} \cite{mM0=PLB}
\begin{eqnarray}\label{M0:DX=gP}
D_{+q}{\bf X}{}^i= 4i\gamma^i_{qp}\Psi_{q}\;  \qquad
\end{eqnarray}
where the anti-hermitean fermionic spinor superfield $\Psi_{q}:= \Psi_{+++ q} $ with SO(1,1) weight 3 is related to the hermitean fermionic field strength in (\ref{M0:G2:=}) by $\Psi_{q}=iG_{+q\; ++}$.

As far as the SYM model is defined on the superspace ${\cal W}^{(1|16)}$ obeying the superembedding equation (\ref{SembEq=M0}), its geometry is characterized by Eqs. (\ref{M0:De+q=}), (\ref{M0:Gij=onshell})  and (\ref{dOm0=0}). This implies that
\begin{widetext}
\begin{eqnarray}\label{M0:(DfDf)Xi=}
\{ D_{+q}, D_{+p}\} {\bf X}{}^i= 4iD_{\#} {\bf X}{}^i \gamma^i_{qp} - i [ {\bf X}{}^i, {\bf X}{}^j] \gamma^j_{qp} + {4i\over 3} {\bf X}{}^j \hat{F}_{\# k_1k_2k_3} \left( \delta^{i[k_1} \gamma_{pq}^{k_2}\delta^{k_3] j} - {1\over 12}\gamma_{pq}^{ijk_1k_2k_3} \right)\;  \qquad
\end{eqnarray}
Using this anticommutation relation together with the superembedding--like equation (\ref{M0:DX=gP}) we find
\begin{eqnarray}\label{M0:Dpsiq=}
& D_{+p}\Psi_{q}\; = {1\over 2}\gamma^i_{pq}D_{\#}{\bf X}^i +
{1\over 16} \gamma^{ij}_{pq} \; [{\bf X}^i, {\bf X}^j]  - {1\over 12} {\bf
X}^i\hat{F}_{\# jkl}\left(\delta^{i[j}\gamma^{kl]}+ {1\over 6}
\gamma^{ijkl}\right)_{pq}
 \; .
\end{eqnarray}
This equation shows that the set of physical fields of the $d=1$, ${\cal N}=16$ SYM model
defined by constraints (\ref{M0:G=sX}) is exhausted by the leading
component of the bosonic superfield ${\bf X}{}^i $, providing the
non-Abelian, $N\times N$ matrix generalization of the Goldstone
field describing a single M$0$-brane in static gauge, and by its
superpartner, the leading component of the fermionic superfield
$\Psi_{q}$ in (\ref{M0:DX=gP}), providing the non-Abelian, $N\times
N$ matrix generalization of the fermionic Goldstone fields describing
a single M0-brane. These can be extracted from the fermionic
coordinate functions of a single M0-brane by fixing the gauge with respect to
local fermionic $\kappa$--symmetry. Notice that in our approach no non-Abelian counterpart of the $\kappa$--symmetry is needed as far as the relative motion of the mM$0$ constituents is described by
matrix counterpart of the physical Goldstone fields of a single brane rather then of the coordinate functions.

This also explains a specific way of realizing the manifest $SO(1,10)$ Lorentz symmetry in our model.
The physical fields of a single brane model are usually extracted by fixing a Lorentz non-covariant gauge (with respect to $\kappa$--symmetry and reparametrization symmetry) and, as a result,  carry the indices of a subgroup of the SO(1,10) Lorentz group, including $SO(9)\times SO(1,1)$ in the M$0$ case. Then our matrix valued fields, being a counterpart of these physical fields, carry the $SO(9)$ indices and definite $SO(1,1)$ waits, while are inert under the $SO(1,10)$ Lorentz group which acts nontrivially on the variables describing the center of energy motion only.

\subsection{Equations of motion and polarization of multiple M0 by flux.}
Next stage is to study the selfconsistency condition of Eq. (\ref{M0:Dpsiq=}). Using the fermionic covariant derivative algebra we can present that in the form
\begin{eqnarray}\label{M0:(DfDf)Psi=}
& \{ D_{+q}, D_{+p}\} {\Psi}_r= 4iD_{\#} {\Psi}_r \delta_{qp} - i [ {\Psi}_r , {\bf X}{}^j] \gamma^j_{qp} + {i\over 3} \hat{F}_{\# ijk} {\Psi}_s \left( \gamma_{qp}^{[i} \gamma_{sr}^{jk]}
 + {1\over 12}\gamma_{pq}^{ijkk_1k_2} \gamma_{sr}^{k_1k_2} \right)\;  \qquad \nonumber \\
 & = D_{+(q}\left(\gamma^iD_{\#}{\bf X}^i +
{1\over 8} \gamma^{ij} \; [{\bf X}^i, {\bf X}^j]  - {1\over 6} {\bf
X}^i\hat{F}_{\# jkl}\left(\delta^{i[j}\gamma^{kl]}+ {1\over 6}
\gamma^{ijkl}\right)\right){}_{p)r} \; . \qquad
\end{eqnarray}
Then, using Eq. (\ref{DFijk=}) and
\begin{eqnarray}
 \label{M0:(DbDf)X=}
 {} [D_{+p} , D_{\#}] {\bf X}^i= i[ {\bf X}^i, {\Psi}_p]  + {i\over 18} F_{\# jkl} (\gamma^{jkl} \gamma^{i})_{pq} {\Psi}_q - i
\hat{T}_{\# [i| \, +q} \gamma_{|j]qp} {\bf X}^j  \; , \qquad
\end{eqnarray}
we find, after some algebra, that the $pq$--trace part of Eq. (\ref{M0:(DfDf)Psi=}) results in the  interacting dynamical equation for the 16-plet of fermionic matrix (super)fields
\begin{eqnarray}\label{M0:DtPsi=}
& D_{\#}\Psi_{q}=- {1\over 4} \gamma^i_{qp} \left[ {\bf X}^i\, , \,
\Psi_{p} \right]   + {1\over 24}  \hat{F}_{\# ijk} \gamma^{ijk}_{qr}\Psi_{r}
 - {1\over 4}  {\bf X}^i\hat{T}_{\# i \, +q}\, .
\end{eqnarray}
We have simplified the final form of the fermionic equation (\ref{M0:DtPsi=}) using the consequence (\ref{hatRS=}) of the supergravity Rarita--Schwinger equation. Using this equation one can also check that the other irreducible parts of the selfconsistency condition (\ref{M0:(DfDf)Psi=}) are satisfied identically (the fact which can be considered as a nontrivial consistency check for our basic equations).

As usual in supersymmetric theories, the higher components in decomposition of the superfield version
of the fermionic equations over the Grassmann coordinates of superspace gives the bosonic equations of motion.  In the case of our multiple M$0$ system, using the commutation relations
\begin{eqnarray}\label{M0:(DbDf)Psi=}
[D_{+p}, D_{\#}] {\Psi}_q=  - i \{ {\Psi}_q, {\Psi}_p\}  + {1\over 72} F_{\# ijk} \gamma^{jkl}_{pr} D_{+r} {\Psi}_q -  {i\over 4} {\Psi}_s \gamma_{sq}^{ij} \gamma_{pr}^{j} \hat{T}_{\# i \, +r}
\;  , \qquad \end{eqnarray}
as well as Eqs. (\ref{M0:(DbDf)X=}), (\ref{DFijk=}) and (\ref{DpT--iq=}), we find
the Gauss constraint
\begin{eqnarray}\label{M0:XiDXi=}
& \left[ {\bf X}^i\, , \, D_{\#}{\bf X}^i\,  \right] = 4i \left\{\Psi_{q}\, , \, \Psi_{q} \right\}  \;  \qquad
\end{eqnarray}
and proper bosonic equation of motion
\begin{eqnarray}\label{M0:DDXi=}
& D_{\#}D_{\#} {\bf X}^i =
{1\over 16} \left[ {\bf X}^j\, , \, \left[ {\bf X}^j\, , \, {\bf X}^i\,  \right]\right]+ i\gamma^i_{qp} \left\{\Psi_{q}\, , \, \Psi_{p} \right\} +  {1\over 4}  {\bf X}^j \hat{R}_{\# j\; \# i} +
{1\over 8} \hat{F}_{\# ijk} \left[ {\bf X}^j\, , \, {\bf X}^k\,  \right] -2i \Psi_{q}\hat{T}_{\# i \, +q}  \; . \qquad
\end{eqnarray}
\end{widetext}
Notice that the Gauss constraint comes from the trace ($\propto \delta_{qp}$) part of the equation
$D_{+p}D_{\#}\Psi_q= [D_{+p}, D_\# ] \Psi_q + D_{\#} (D_{+p}\Psi_q)$ written with the use of Eqs. (\ref{M0:Dpsiq=}) and (\ref{M0:DtPsi=}). The second order equations of motion (\ref{M0:DtPsi=}) is obtained from the  $\propto \gamma^i_{qp}$ irreducible part of that equation, while the other irreducible parts ($\propto \gamma^{ij}_{qp}$, $\propto \gamma^{ijk}_{qp}$ and $\propto \gamma^{ijkl}_{qp}$) are satisfied identically when the consequence (\ref{hatRS=}) of the supergravity Rarita-Schwinger equation is taken into account. This provides one more consistency check of our approach.

The bosonic equation (\ref{M0:DDXi=}) has an interesting structure, particularly in its part describing coupling to the generic supergravity background. The fourth term in the {\it r.h.s.} of this equation, $ \hat{F}_{\# ijk} \left[ {\bf X}^j\, , \, {\bf X}^k\,  \right]$, is typical for `dielectric coupling' characteristic for the Emparan-Myers `dielectric brane effect' \cite{Emparan:1997rt}, \cite{Myers:1999ps}. It is essentially non-Abelian as far as in the Abelian case this contribution  vanishes. This is the case also for the first and the second terms in the {\it r.h.s.} of  (\ref{M0:DDXi=}), which are also present in the case of flat background without fluxes and  in 1d dimensional reduction of 10D SYM (which is clearly not the case for the other three terms describing interactions with fluxes of 11D supergravity).

The third term in the {\it r.h.s.} (\ref{M0:DDXi=}) is linear in ${\bf X}^j$ and thus give rise to a mass term  for this $su(N)$ valued matrix bosonic (super)field. The corresponding mass matrix is induced by fluxes: namely, it is  expressed through the pull--back of the specific projection of Riemann tensor, $\hat{R}_{\#i\; j\#}$ of Eq. (\ref{M0:RFlux}). Notice that this latter is symmetric in its SO(9) vector indices (\ref{hatR[]=0}) which is important because otherwise  equation  (\ref{M0:DDXi=}) would look  non-Lagrangian. 
Notice also that, due to the consequence (\ref{hatEiEq=singlet}) of the supergravity Einstein equation, when the multiple M$0$ system interacts nontrivially with the four form fluxes, the field ${\bf X}^j$ , describing the relative motion of the mM$0$ constituents, is always massive as far as the trace of its mass matrix is nonvanishing.

\section{BPS equations and supersymmetric bosonic solutions of the multiple M$0$ equations}
\setcounter{equation}{0}

\subsection{Supersymmetry preservation by single M$0$ }

The presence of M$0$-brane breaks one-half of the spacetime supersymmetry. This can be easily seen from Eqs.  (\ref{Eua=e++u--}) and  (\ref{Ef=efv}) describing the on-shell superembedding of the M$0$ worldline superspace or of the center of energy superspace of mM$0$ system, ${\cal W}^{(1|16)}$, into the target 11D superspace ${\Sigma}^{(11|32)}$. Indeed, in superspace formulation the local supersymmetry transformations of a supergravity model can be identified with supertranslations  in the fermionic directions,
 \begin{eqnarray}\label{susy=11D}
 \varepsilon^\alpha =  \delta Z^M E_M^\alpha ({Z})
=: i_\delta E^\alpha\; . \qquad
\end{eqnarray}
Then Eq. (\ref{Ef=efv}) implies that, if 32-component spinor parameter $\varepsilon{}^{\alpha}$ describes a supersymmetry preserved by some M$0$--brane, its  pull--back $\hat{\varepsilon}{}^{\alpha}$  to ${\cal W}^{(1|16)}$ is expressed through $16$ parameters \begin{eqnarray}\label{susy=1d}
 \epsilon^{+q}=\delta \zeta^{\cal M} e_{\cal M}^{+q}(\zeta)=: i_\delta e^{+q}  \;  \qquad
\end{eqnarray}
of the local worldline supersymmetry,
\begin{eqnarray}\label{susy=1/2susy}
& \hat{\varepsilon}{}^{\alpha}= \epsilon^{+q} v^{-\alpha}_q
 \; . \qquad
\end{eqnarray}

Thus, in a completely supersymmetric background a solution of   M$0$-brane equations can preserve 16 or less of 32 target (super)space supersymmetries. If the supergravity background preserves a part of supersymmetries, the situation becomes more complicated as the number of preserved supersymmetries may become dependent on  details of M$0$-brane motion (see recent \cite{Dima+Linus+09} for the specific case of strings and branes in type IIA superspace describing 3/4 supersymmetric $AdS_4\times {CP}^3$ background).

The superembedding approach allows to make some general statements about supersymmetry preservation by M$0$-brane motion in a purely bosonic background. In superspace such backgrounds are characterized by \begin{eqnarray}\label{Tbbf=0} T_{ab}{}^\alpha (x)=0
 \; . \qquad
\end{eqnarray}
Then the pull--back of this fermionic field strength to the worldvolume and its projections also vanish. Taking into account that only the projection (\ref{M0:fFluxe}) enters the description of the M$0$ worldline superspace geometry and, through that, its dynamics, one sees that the part of
worldline supersymmetry (part of the one half of the target space supersymmetry) preserved by a certain M$0$ motion is characterized by parameters which obey
\begin{eqnarray}\label{susyDT=0}
& \epsilon^{+p}\, (D_{+p} \hat{T}_{\# \, i\, + q})\vert_0= 0 \; \;  \qquad
\end{eqnarray}
with $\vert_0:=\vert_{\eta^p=0}$.

In our approach this equation also determines the ${\cal W}{}^{(1|16)}$ supersymmetry (a part of the one half of the target space supersymmetry) preserved by the center of energy motion of the mM$0$ system.
But in this case this is not the end of story
as the supersymmetry preserved by the center of
energy motion can be either preserved or broken by the relative
motion of the mM$0$ constituents.

\subsection{Supersymmetry preservation by multiple mM$0$ system}

The supersymmetry transformation $\delta_{susy}\psi_q(\tau)$  of the $N\times N$ matrix fermionic
field $\psi_q(\tau):=\Psi_q(\tau, 0)\equiv \Psi_q \vert_{0}$
can be identified as
$\delta_{susy}\psi_q(\tau)=\epsilon^{+p}D_{+p}\Psi_q\vert_{0}$.
Then the preservation of supersymmetry for bosonic solutions of the equations describing relative motion of mM$0$ implies
\begin{eqnarray}\label{susyDPsi=0}
 \epsilon^{+p}(D_{+p}\Psi_q )\vert_{0} =0
 \; . \qquad
\end{eqnarray}

Furthermore, using Eqs. (\ref{DpT--iq=}), (\ref{FFS=}) and (\ref{M0:Dpsiq=}) one can present the system of equations
(\ref{susyDT=0}) and (\ref{susyDPsi=0}) for the parameter of supersymmetry preserved by the mM$0$ system in the following from
\begin{widetext}
\begin{eqnarray}\label{M0:epN=BPS}
 \epsilon^{+p} {\mathcal N}_{i\; pq}=0\; , & \qquad   {\cal N}_{i\; pq}:=  D_{+p} \hat{T}_{\# \, i\, + q} =
   {1\over 2} \gamma^j_{pq} \left( \hat{R}_{\# ij
\#} - {1\over 6}   \hat{F}_{\# i k_1k_2} \hat{F}_{\# j k_1k_2}-{1\over 54}\delta^{ij} (\hat{F}_{\# k_1k_2k_3})^2\right)+ \nonumber \\ & + {1\over 3} D_{\#} \hat{F}_{\# ijk} \left(\delta^{i[j}\gamma^{kl]}_{pq} + {1\over 6}
\gamma^{ijkl}_{pq} \right) +{1\over 9} \gamma^{j_1j_2j_3}_{pq}  \hat{F}_{\# i j_1k} \hat{F}_{\# k j_2j_3} + \qquad \nonumber \\ &  + {1\over 72} \gamma^{ k_1 k_2 k_3 k_4 k_5}_{pq} \left( \hat{F}_{\# i k_1k_2} \hat{F}_{\# k_3k_4k_5} +\delta^{i}_{[k_1} \hat{F}_{\# k_2k_3|j}\hat{F}_{\# j|k_4k_5]}\right)
  \; , \qquad  \\
\label{M0:epM=BPS}
 \epsilon^{+p} {\bf M}_{pq}=0\; , & \qquad
{\bf M}_{pq}:= D_{+p}\Psi_q =
\left(\gamma^i_{pq}D_{\#}{\bf X}^i +
{1\over 8} \gamma^{ij}_{pq} \; [{\bf X}^i, {\bf X}^j]  - {1\over 6} {\bf
X}^i\hat{F}_{\# jkl}\left(\delta^{i[j}\gamma^{kl]}+ {1\over 6}
\gamma^{ijkl}\right)_{pq}\right)
 \; . \qquad
\end{eqnarray}
\end{widetext}
Here and below  we denote the leading component of superfield by the same symbol as the whole superfield, {\it i.e.}, if treating equations in terms of superfield, we assume $\vert_0$ ($:=\vert_{\eta^p=0}$) symbol, but do not write this explicitly.

\subsection{1/2 BPS equations for single M$0$--brane }

It is natural to begin with the study of 1/2 BPS equations for the more conventional case of single M$0$--brane. This  preserves  one-half of the target space supersymmetry if the equation (\ref{M0:epN=BPS}) is satisfied for arbitrary SO(9) spinor $\epsilon^{+p}$. Hence the 1/2 BPS equations for single M$0$--brane are enclosed in the equation
\begin{eqnarray}\label{DpT--iq=0}
 \mathcal{N}_{i\; pq}= 0 \; , \qquad
\end{eqnarray}
where  $\mathcal{N}_{i\; pq}$ is defined in (\ref{M0:epN=BPS}). Decomposing this on the irreducible parts,  one finds the following set of the 1/2 BPS equations for single M$0$--brane
\begin{eqnarray} \label{DpTq=0->R}
 \hat{R}_{\# ij\#} - {1\over 6}\hat{F}_{\# ikl}\hat{F}_{\# klj}- {1\over 54}\delta^{ij} (\hat{F}_{\# k_1k_2k_3})^2=0 \; , \\ \label{DpTq=0->DF}
 D_{\#} \hat{F}_{\# ijk} =0 \; , \qquad  \\
 \label{DpTq=0->FF3}
 \hat{F}_{\# ij[k_1}\hat{F}_{\# k_2k_3]j}=0\; ,\qquad  \end{eqnarray} \begin{eqnarray}
 \label{DpTq=0->FF4}
 \hat{F}_{\# j[k_1k_2}\hat{F}_{\# k_3k_4]j}=0\; ,\qquad  \\
 \label{DpTq=0->FF5}
 \hat{F}_{\# i[k_1k_2}\hat{F}_{\# k_3k_4k_5]}=0\; .\qquad
\end{eqnarray}
Notice that Eq. (\ref{DpTq=0->R}) cannot be obtained from pull--back of the Einstein equation of supergravity, Eq. (\ref{Eq=Einstein}), but its trace coincides with the consequence (\ref{hatEiEq=singlet}) of this Einstein equation.

Eq. (\ref{DpTq=0->DF}) implies that the pull--back of the four form flux is essentially constant (independent on the proper time coordinate; notice that one can fix the gauge $A_\#=0$). As far as the algebraic equations (\ref{DpTq=0->FF3})--(\ref{DpTq=0->FF5}) are concerned, they are solved by
\begin{eqnarray}\label{F=F3-1/2BPS}
 \hat{F}_{\# ijk}= 3/4  w^i_I w^j_J w_K^k \epsilon^{IJK} \; , \qquad \begin{cases}i=1,\ldots, 9 \cr
 I=1,2,3\; \end{cases}
\end{eqnarray}
where  $\epsilon^{IJK}=\epsilon^{[IJK]}$ is the Levi-Civita symbol, $\epsilon^{123}=1$, and  $9\times 3$ matrices $w^i_I$ obey $D_{\#} w^i_I =0$.

Thus, a certain M$0$ motion can preserve 1/2 of 32 target spacetime supersymmetries if the projection of the pull--back of the target superspace flux to the M$0$ worldline obeys (\ref{F=F3-1/2BPS}).


\subsection{1/2 BPS equations for multiple M$0$-brane  system}

In our approach, a certain configuration of multiple M$0$ system can preserve 1/2 of the complete supersymmetry if the center of energy motion obeys Eq. (\ref{DpT--iq=0}) and the relative motion of  mM$0$ constituents preserves all the 16 supersymmetries preserved by the center of energy motion. Thus we have to assume that  (\ref{M0:epM=BPS}) is obeyed for an arbitrary $\epsilon^{+p}$ so that the system of 1/2 BPS equations
for the multiple M$0$ system includes Eq. (\ref{DpT--iq=0}),
\begin{widetext}
 \begin{eqnarray}\label{M0:eBPSb=}
& {\cal N}_{i\; pq} =
   {1\over 2} \gamma^j_{pq} \left( \hat{R}_{\# ij
\#} - {1\over 6}   \hat{F}_{\# i k_1k_2} \hat{F}_{\# j k_1k_2}-{1\over 54}\delta^{ij} (\hat{F}_{\# k_1k_2k_3})^2\right)+ {1\over 3} D_{\#} \hat{F}_{\# ijk} \left(\delta^{i[j}\gamma^{kl]}_{pq} + {1\over 6}
\gamma^{ijkl}_{pq} \right) + \nonumber \\ & + {1\over 9} \gamma^{j_1j_2j_3}_{pq}  \hat{F}_{\# i j_1k} \hat{F}_{\# k j_2j_3} + {1\over 72} \gamma^{ k_1 k_2 k_3 k_4 k_5}_{pq} \left( \hat{F}_{\# i k_1k_2} \hat{F}_{\# k_3k_4k_5} +\delta^{i}_{[k_1} \hat{F}_{\# k_2k_3|j}\hat{F}_{\# j|k_4k_5]}\right) =0
  \; , \qquad
\end{eqnarray}
and
\begin{eqnarray}\label{M0:eBPS=}
 {\bf M}_{pq}:= \gamma^i_{pq}D_{\#}{\bf X}^i +
{1\over 8} \gamma^{ij}_{pq} \left( [{\bf X}^i, {\bf X}^j]  - {4\over 3}
F_{\# ijk}{\bf X}^k \right)
+ {1\over 36} {\bf X}^iF_{\# jkl}
\gamma^{ijkl}_{pq}=0
 \; . \qquad
\end{eqnarray}
 \end{widetext}
Furthermore, as the 1/2 BPS equations for the  center of energy motion (\ref{M0:eBPSb=}) ((\ref{DpT--iq=0})) is equivalent to the set of equations (\ref{DpTq=0->R}) and (\ref{F=F3-1/2BPS}), one can search for 1/2 BPS solutions of multiple M$0$ system on the basis of Eq. (\ref{M0:eBPS=}) with the projection of the pull--back of the 4-form flux defined by Eq. (\ref{F=F3-1/2BPS}) with $D_\# w_I^i=0$.

Actually, it is instructive to make a step back and do not use (\ref{F=F3-1/2BPS}) from the very beginning.  Decomposing Eq. (\ref{M0:eBPS=}) on the irreducible parts, we find
\begin{eqnarray}\label{M0:BPS1=} D_{\#}{\bf X}^i=0 \; , \qquad \\
\label{M0:BPS2=}
\left[ {\bf X}^i\, , \,
{\bf X}^j \right]={4\over 3} \hat{F}_{\# ijk}   {\bf X}^k  \; , \qquad \\ \label{M0:BPS3=}
 {\bf X}^{[i}\,  \hat{F}_{\#}{}^{jkl]}=0  \; . \qquad
\end{eqnarray}
One dimensional gauge connection is always trivial, and so is the
bosonic part $e^\# A_\#$ of our $d=1$ ${\cal N}=16$ superspace
connection $A=e^\# A_\#+ e^{+q} A_{+q}$. Hence Eq. (\ref{M0:BPS1=}) means that for the 1/2 BPS configuration of mM$0$ the relative motion of the mM$0$ constituents is descried by essentially constant $N\times N$ matrices
obeying Eqs. (\ref{M0:BPS2=}) and (\ref{M0:BPS3=}). Now one notice that, as far as pull--back of the flux is a number while our ${\bf X}^{i}$ are $N\times N$ matrices, the solutions of Eq. (\ref{M0:BPS3=}) have a nontrivial matrix structure only if the flux has the form   $\hat{F}_{\# ijk}= 3/4 w^i_I w^j_J w_K^k \epsilon^{IJK}$, as in Eq. (\ref{F=F3-1/2BPS}) dictated by the supersymmetry preservation by the center of energy motion. Eqs. (\ref{M0:BPS2=}) and (\ref{M0:BPS3=}) with such a flux are solved by the following fuzzy 2--sphere configuration
\begin{eqnarray}\label{M0:BPS=fuS3}
 {\bf X}^i = w^i_I T^I\; , \qquad \;
 {}[  T^I , T^J ] = \epsilon^{IJK} T^K \; . \qquad
\end{eqnarray}
Here the triplet of $N\times N$ matrices $T^I$ provides $N\times N$ representation of the $SU(2)$ generators. As in (\ref{F=F3-1/2BPS}),  $w^i_I$ are $9\times 3$ matrices ($i=1,...,9$, $I=1,2,3$) playing the r\^ole of bridge between representations of the $SO(9)$ and $SO(3)=SU(2)$ groups.

A simple particular case of the above 1/2 BPS solution is the one with $w^i_I=f \delta_i^I$, which occurs when the projection of the flux has only one nonvanishing basic component $\hat{F}_{\# 123}$ and the relative positions of mM$0$ constituents are described by the set of only three nonvanishing bosonic $N\times N$ matrices
${\bf X}^i =  (fT^1, fT^2, fT^3, 0,0,0,0,0,0)$,
\begin{eqnarray}
\label{M0:BPS=fuS3'}
 && {\bf X}^i = f\delta_I^iT^I\; , \qquad
 {}[  T^I , T^J ] = \epsilon^{IJK} T^K
 \; . \qquad \\
\label{M0:fluxBPS=123}
  && \hat{F}_{\# IJK}=3/4f^3\epsilon^{IJK} \; , \qquad \nonumber \\
  && \hat{F}_{\# ijk}=0 \quad for (i,j,k)\not= permutation \;  of\; 123 \;  \qquad
\end{eqnarray}

As we have already stated, the configuration (\ref{M0:BPS=fuS3'}) is called fuzzy two sphere \cite{Hoppe82+}.
Similar configuration  was shown to solve the purely bosonic equations for 'dielectric D$0$-branes' following
from the $p=0$ Myers action for a particular type IIA background \cite{Myers:1999ps}. In that case three nonvanishing $N\times N$ matrices ${\bf X}^I = fT^I$ define the set of eight $u(N)$ valued matrices ${\bf X}^{\hat{i}} = (fT^I,0,0,0,0,0)$ describing both the relative motion and the motion of the center of mass of the purely bosonic Myers D$0$-branes. The issue of supersymmetry was not addressed  in  \cite{Myers:1999ps} as far as the supersymmetric generalization of the Myers action was not known.

In contrast, the fuzzy sphere solution of our multiple M$0$ equations is supersymmetric by construction and can be considered as modeling M$2$--brane. Moreover, our approach allows to see  explicitly the origin of the $SU(2)$ structure constant in the four form flux, Eq.  (\ref{M0:fluxBPS=123}), and that this form is essentially the only nonvanishing flux allowed by the conditions of preservation of 1/2 of the target space supersymmetry.

\subsection{A particular class of  1/4 BPS states and Nahm equation}

The set of 1/4 BPS states of the mM$0$ system is split on different sectors. Indeed, for this case $8$ of $16$ supersymmetries which might be preserved by embedding of superspace ${\cal W}^{(1|16)}$ into $\Sigma^{(11|32)}$ are broken and, in generic case, some number $r$ of them are broken by center of mass motion and $(8-r)$ --  by the relative motion of the M$0$ constituents ($r \; \leq 8$ in the context of our 1/4 BPS discussion). In other words $r$ of the 16 components of $SO(9)$ spinor $\epsilon^p$  can be set to zero by Eq. (\ref{M0:epN=BPS}), corresponding to center of energy motion,  and the remaining  part - $(8-r)$ - by Eq.  (\ref{M0:epM=BPS}), characterizing the relative motion of mM$0$ constituents. Here we will restrict ourselves by considering one specific sector, with 1/2 supersymmetric center of energy `motion' and additional 1/4 of supersymmetry broken by the relative motion of mM$0$ constituents.

In the assumption that the center of mass `motion' obeys the 1/2 BPS condition (\ref{DpT--iq=0}), the BPS states preserving 1/4 of target space supersymmetry should obey
\begin{eqnarray}\label{M0:1/4BPS=P}
& {\cal P}_{pr} {\bf M}_{rq}={1\over 2} (1-\bar{\gamma})_{pr}{\bf M}_{rq}=0\; .
\qquad
\end{eqnarray}
where the $16\times 16$ matrices ${\bf M}_{pq}$ is defined in Eq. (\ref{M0:epM=BPS}) and ${\cal P}_{pq}={1\over 2}(1-\bar{\gamma})_{pq}$ is the rank 8 projector, ${\cal P}{\cal P} ={\cal P}$, constructed from
the  matrix $\bar{\gamma}_{pq}$ which obeys
\begin{eqnarray}\label{M0:1/2BPS=}
\bar{\gamma}^2=I\; , \qquad tr(\bar{\gamma}):=\bar{\gamma}_{qq}=0\; . \qquad
\end{eqnarray}

We are going to show now that the famous  Nahm equation \cite{NahmEq} appears as a particular $SO(3)$ invariant case of Eq. (\ref{M0:1/4BPS=P}).

Let us set $\hat{F}_{\# ijk}=0$, consider bosonic solutions with only three nonvanishing $N\times N$ matrices of nine,
\begin{eqnarray}\label{M0:Xi=XI3}
{\bf X}^i = ({\bf X}^I, \underbrace{0,...,0}_6)=({\bf X}^1,{\bf X}^2, {\bf X}^3, 0,0,0,0,0,0)\; ,
\qquad
\end{eqnarray}
and identify  $\bar{\gamma}= i \gamma^{123}$. Then  $\bar{\gamma}\gamma^{IJ}=-i\epsilon^{IJK}\gamma^K$ and Eq. (\ref{M0:1/2BPS=}) reads
\begin{eqnarray}\label{M0:1/2BPS=Nahm}
((1-\bar{\gamma})\gamma^I)_{pq} \left( D_{\#}{\bf X}^I +
{i\over 8} \epsilon^{IJK}  [{\bf X}^J, {\bf X}^K]\right) =0\; .
\qquad
\end{eqnarray}
This implies the Nahm equation \cite{NahmEq}
\begin{eqnarray}\label{M0:Nahm=1/2}
 D_\# {{\bf X}}{}^I +
{i\over 8} \epsilon^{IJK}  [{\bf X}^J, {\bf X}^K] =0\; .
\qquad
\end{eqnarray}
The literal coincidence with the original form of Nahm equation appears when we fix the gauge $A_\#=0$ and set to zero the normal bundle connection, so that $D_\# {{\bf X}}{}^I= \dot{{\bf X}}{}^I:= {\partial {\bf X}^I\over \partial\tau}$.

Thus the famous Nahm equation which has a fuzzy--sphere--related fuzzy funnel solution appears as a particular case of SO(3) symmetric 1/4 BPS equation for our multiple M$0$ system in the background with vanishing 4-form flux. Hence, surprisingly enough, the origin of the Levi-Civita symbol $\epsilon^{IJK}$ in the Nahm equation for mM$0$ system is not the 11D supergravity flux, as one might expect, but rather the requirement of $SO(3)$ symmetry of the particular 1/4 BPS configurations described by three nonvanishing components of ${\bf X}^i$.

\bigskip

\section{Conclusion and discussion}
\setcounter{equation}{0}

In this paper we have obtained equations of motion for the systems of multiple M$0$-branes (multiple M-waves or mM$0$-system) in an arbitrary supergravity background. These equations are derived in the frame of superembedding approach, defining 1d SYM connection restricted by the set of constraints on the 1d ${\cal N}=16$ superspace ${\cal W}^{(1|16)}$ the embedding of which in the generic 11D supergravity superspace $\Sigma^{(11|32)}$ is determined by the superembedding equation. The same superembedding equation defines the embedding of the worldvolume superspace of a single M$0$-brane, which is the massless 11D superparticle, and encodes its equations of motion which imply that the M$0$ worldline is light--like. In the case of mM$0$ the superembedding equations for the superspace ${\cal W}^{(1|16)}$ results in the equations describing the center of energy motion of mM$0$ system, which is also characterized by a light-like geodesic.

The equations for the relative motion of mM$0$ constituents follow from the constraints imposed on the field strength of the SYM connection. These, together with the center of energy equations of motion, provide a generalization of the Matrix model \cite{Banks:1996vh} for an arbitrary supergravity background. Notice that Matrix model have been known before only for a very few particular backgrounds, including the maximally  supersymmetric pp-wave background \cite{BMN}.
Hence the natural application of the present approach is to use our general equations to obtain Matrix model in physically interesting backgrounds. In particular the equations for Matrix model in $AdS_4\times S^7$ and $AdS_7\times S^4$ backgrounds can be straightforwardly obtained in this manner.

This will be the subject of our subsequent study which will also include the derivation of our mM0 equations in  supersymmetric pp-wave background and comparison of  the result with the BMN (Berenstein--Maldacena--Nastase) matrix model \cite{BMN}. In this respect we should mention that there exist some conjectures \cite{SheikhJabbari:2004ik} that the BMN model actually provides the description of the M(atrix) theory in an arbitrary background.  The only comment we would like to make in this respect in the present paper is that, even if such a conjecture were proved to be correct, it would be a nontrivial problem to extract the information on a certain system in definite non-pp-wave background from it
\footnote{In particular, as argued in \cite{Feinstein:2002ay}, even the $AdS\times S$ background is completely determined by the {\it two} orthogonal Penrose limits: {''Having only one limit does not determine the whole spacetime. Thus, the two orthogonal Penrose limits form a sort of classical holographic
boundary for the background with D - 2 commuting Killing directions.''}\cite{Feinstein:2002ay}.}
so that, in our view,  it would be certainly useful anyway to have an explicit form of the M(atrix) model in an arbitrary supergravity background.

We should also notice that all the terms with background contributions to the {\it r.h.s.}'s of our mM$0$ equations are linear in fluxes which is in disagreement with expectations based on the study of the Myers-type actions \cite{Myers:1999ps,YLozano+=0207}. Although the Myers action is purely bosonic and resisted all the attempts of its straightforward supersymmetric and Lorentz covariant generalization eleven years (except for the cases of lower dimensional and lower co-dimensional Dp-branes \cite{Dima01}), taking into account a particular progress in this direction reached recently in the frame of the boundary fermion approach \cite{Howe+Linstrom+Linus} and also evidences from the string amplitude calculations, we should not exclude the possibility that the above mentioned discrepancy implies that our approach gives only an approximate description of the Matrix model interaction with supergravity fluxes.

If this is the case, a way to search for a more general interaction lays through modification of the basic equations of our superembedding approach, namely the superembedding equation, defining the embedding of the mM$0$ center of energy superspace ${\cal W}^{(1|16)}$ into the target $D=11$ supergravity superspace $\Sigma^{(11|32)}$, and  the basic constraints of the $d=1$, ${\cal N}=16$ SYM model on the center of energy superspace. (Notice that the possible modification of the basic superembedding-type equations in the boundary fermion approach was suggested recently  in \cite{Howe+Linstrom+Linus=2010}). The problem of the deformation of the basic constraint determining the equations for the relative motion of mM$0$ constituents is the curved superspace generalization of the studies in \cite{Cederwall:2001td,Movshev:2009ba,Howe+Linstrom+Linus=2010}. However, unfortunately, if we allow consistent deformations of the basic equations of our superembedding approach, although  most probably these exist, they would certainly make the equations very complicated up to being unpractical.

Interestingly enough, if we do not deform the superembedding equation, but allow for a deformation of the
$d=1$, ${\cal N}=16$ $SU(N)$ SYM constraints on the center of mass superspace (see \cite{mM0=PLB,mM0=PRL}), the situation seems to be much more under control due to the rigid structure of the mM$0$ equations \cite{mM0=PRL}. In this case the center of mass motion and supersymmetry of the corresponding superspace is influenced only by the projections (\ref{M0:fFluxe}) of the supergravity fluxes, so that it is reasonable to assume that only these fluxes can enter the equations of  relative motion of the mM$0$ constituents. Then,
the requirement of SO(1,1)$\times$SO(9) symmetry leaves very few possibilities to add the new terms to the ones already present in the {\it r.h.s.}'s of the equations (\ref{M0:DtPsi=})--(\ref{M0:DDXi=}) \cite{mM0=PRL}. In particular, the only possible nonlinear contribution to the {\it r.h.s.} of the bosonic equations (\ref{M0:DDXi=}) is ${\bf X}^j \hat{F}_{\# jkl} \hat{F}_{\# ikl} $ describing the contribution  proportional to the second power of the 4-form flux to the mass matrix of the $su(N)$-valued fields ${\bf X}^j $. \footnote{One might also propose the term ${\bf X}^i \hat{F}_{\# jkl} \hat{F}_{\# jkl}\equiv {\bf X}^i (\hat{F}_{\# jkl})^2$, but this reduces  to ${\bf X}^i \hat{R}_{\# j\, \# j}$ due to the consequence (\ref{hatEiEq=singlet})  of the supergravity Einstein equation (\ref{Eq=Einstein}).} To resume, we cannot exclude the possibility that  our approach gives only an approximate description of the Matrix model interaction with supergravity fluxes, but if so, it is Lorentz covariant, supersymmetric and going beyond the $U(N)$ SYM approximation.

We also used superembedding approach to obtain BPS equations for supersymmetric solutions of these mM$0$ equations. As an example we have shown that the 1/2 BPS equations in the presence of 4-form flux have the fuzzy sphere solution modeling M$2$-brane by a 1/2 supersymmetric configuration of multiple M$0$. We also found that the Nahm equation \cite{NahmEq} appears as a particular SO(3) invariant case of the 1/4 BPS equations in the absence of the four form flux.

The further study of our mM$0$ BPS equations and search for new solutions of their equations of motion is an interesting problem for future study. A particularly intriguing problem is to search for a description (better to say, modeling) of the M5 brane and/or M2-M5 system in this framework. The popular candidate for the description of this latter is the Basu-Harvey equation \cite{BasuHarvey}
\begin{eqnarray}\label{M0:1/2BPS=5a1}
 D_{\#}{\bf X}^{\tilde{I}} =  \epsilon^{5\tilde{I}JKL} {\bf X}^5 {\bf X}^J{\bf X}^K{\bf X}^L  \; , \qquad \tilde{I}=1,...,4\; . \quad \nonumber
\end{eqnarray}
We have not succeed in deducing this equation from the BPS conditions for supersymmetric solution of our mM$0$ equations, so that this remains an interesting open problem for future study.

Probably the description of higher branes requires to pass from one dimensional Matrix model to its higher dimensional counterparts, beginning from Matrix string \cite{MatrixString} (see \cite{Nastase-ABJM} for arguments in favor of this). In this respect it is interesting to check a possibility to extend our superembedding approach, as it is developed for mM$0$ and mD$0$, for the case of higher $p$ multiple $p$-brane systems beginning from type IIB multiple D-strings or multiple $(p,q)$--strings. This will be the subject of our future study.

\bigskip
\acknowledgements{
The author is thankful to Dima Sorokin, Yolanda Lozano and Alex Feinstein  for useful discussions and to organizers of the VIII Simons Workshop,  especially to Martin Rocek, for hospitality in Stony Brook on a final stage of this work which was supported in part  by the research grants FIS2008-1980 from the MICINN of Spain and the Basque Government Research Group Grant ITT559-10.}

\newpage

\begin{widetext}

\appendix{{\bf Appendix A: 11D and 9d Gamma matrices}}
\renewcommand{\theequation}{A.\arabic{equation}}
\setcounter{equation}{0}

A convenient  $SO(1,1)\otimes SO(9)$ invariant
representations for the eleven dimensional gamma matrices and charge conjugation matrix read
\begin{eqnarray}\label{11DG=1+9+1}
& (\Gamma^{{a}})_{\underline\alpha}{}^{\underline{\beta}}
 \equiv \left({1\over 2}(\Gamma^{\#} + \Gamma^{=}), \Gamma ^{i}, {1\over 2}(\Gamma^{\#} - \Gamma^{=}) \right)\; , \qquad a=0,1,\ldots,9, 10 \; , \qquad i=1,\ldots, 9 \; , \qquad
\nonumber
\\
 \label{11DG=} & (\Gamma ^{\#})_{\alpha}{}^{\beta}
 = \left( \begin{matrix} 0 & 2i\delta_{pq}
 \cr 0 & 0 \end{matrix}  \right)\;  , \qquad  (\Gamma ^{=})_{\alpha}{}^{\beta}
 = \left( \begin{matrix} 0 & 0
 \cr -2i\delta_{pq} & 0 \end{matrix}  \right)\;  , \qquad  (\Gamma ^{i})_{\alpha}{}^{\beta}
 = \left( \begin{matrix} -i \gamma^{i}_{pq} & 0
 \cr 0 & i\gamma^{i}_{pq} \end{matrix}  \right)\;  , \qquad
 \\
\label{11DC=}
&
C_{{\alpha}{\beta}}  =- C_{{\beta}{\alpha}}=
 \left( \begin{matrix} 0 & i\delta_{pq}
 \cr -i\delta_{pq} & 0
\cr\end{matrix} \right)=  (C^{-1}){}^{{\alpha}{\beta}}=:  C^{{\alpha}{\beta}}\;  . \qquad
\end{eqnarray}
These are imaginary as far as we use the mostly minus metric convention so that the flat spacetime metric reads $\eta_{ab}=diag (1,-1,\ldots ,-1)$.

In (\ref{11DG=1+9+1}) $\gamma^{i}_{pq}$ are $16\times 16$ d=9 Dirac matrices. These are symmetric $\gamma^{i}_{pq}=\gamma^{i}_{qp}$, and possesses the following properties
\begin{eqnarray}
 \label{9dgammas}
\gamma^{(i}\gamma^{j)}=  \delta^{ij} I_{16\times 16} \; , \qquad \gamma^{i}_{pq}=\gamma^{i}_{qp}:= \gamma^{i}_{(pq)} \; , \qquad \gamma^{i}_{(pq}\gamma^{i}_{r)s} = \delta_{(pq}\delta_{r)s}
\; . \qquad
\end{eqnarray}
The d=9 charge conjugation matrix is also symmetric, which allows to chose its representation by Kronecker delta symbol $\delta_{qp}$ and do not distinguish upper and lower Spin(9) (SO(9) spinor) indices.
Notice that the matrices $\gamma^{ij}_{qp}$ and $\gamma^{ijk}_{qp}$ are antisymmetric so that the complete basis for the set of $16\times 16 $ symmetric matrices is provided by $\delta_{pq}$, $\gamma^i_{pq}$, $\gamma^{ijkl}_{pq}$,
\begin{eqnarray}
 \label{9d=symg}
 \delta_{r(q} \delta_{p)s}={1\over 16} \delta_{pq} \delta_{rs} + {1\over 16} \gamma^i_{pq} \gamma^i_{rs} + {1\over 16\cdot 4!} \gamma^{ijkl}_{pq} \gamma^{ijkl}_{rs}
\; . \qquad
\end{eqnarray}
In our conventions $\gamma^{123456789}_{qp}= \delta_{qp}\,$ and, consequently,
\begin{eqnarray} \label{9d:g5=g4}
\gamma^{i_1\ldots i_7}_{qp}= -{1\over 2} \epsilon^{i_1\ldots i_7jk} \gamma^{jk}_{qp}\; , \qquad \\
 \label{9d:g5=g4}
\gamma^{i_1\ldots i_5}_{qp}= {1\over 4!} \epsilon^{i_1\ldots i_5j_1\ldots j_4} \gamma^{j_1\ldots j_4}_{qp}\; . \qquad
\end{eqnarray}
This, together with (\ref{11DG=1+9+1})  implies that our 11D dirac matrices obey
\begin{eqnarray}
 \label{11Dgammas}
\Gamma^0\Gamma^1\ldots \Gamma^9\Gamma^{(10)}= {1\over 2} \Gamma^{\#} \Gamma^{=}\Gamma^1\ldots \Gamma^9 = - iI_{32\times 32}\; . \qquad
\end{eqnarray}

\bigskip

\appendix{{\bf Appendix B:  Some properties of moving frame and spinor moving frame variables associated to the massless superparticle. }}
\renewcommand{\theequation}{B.\arabic{equation}}
\setcounter{equation}{0}

Here we collect some useful equations describing properties of moving frame and spinor moving frame variables
(\ref{harmUin}), (\ref{harmVin}).

Moving frame variables appropriate to the description of massless
$D$ dimensional (super)particle were also called light-cone
harmonics in \cite{Sok} and Lorentz harmonics in \cite{B90}. They
are defined  as columns of the $D\times D$ Lorentz group matrix of
which obey the constraints
\begin{eqnarray}\label{harmUinAPP}
U_b^{(a)}= (u_b^{=}, u_b^{\#}, u_b^{i})\;  \in \; SO(1,D-1) \qquad
\Leftrightarrow \qquad
\begin{cases}
U\eta U^T = \eta \quad \Leftrightarrow \delta_a^b= {1\over
2}u_a^{\#}u^{b=} + {1\over 2}u_a^{=}u^{b\# } - u_a^{i}u^{b i} \cr {}
\cr  U^T\eta U = \eta \quad \Leftrightarrow
\begin{cases} u_a^{=}u^{a=}=0 \; , \quad u_a^{\# }u^{a\#}=0 \; , \cr u_a^{=}u^{a\#}=2
\; , \quad u_a^{=}u^{a\, i}=0 \; , u_a^{\#}u^{a\, i}=0 \; , \quad
\cr u_a^{i}u^{a\, j}=- \delta^{ij}
\end{cases}\end{cases} \\ \nonumber b=0,1,\ldots , (D-2), (D-1) \; , \qquad (a)=(\#,=,1,
\ldots , (D-2)) \; .
\end{eqnarray}

The spinor moving frame variables or {\it spinorial harmonics} (see \cite{B90,Ghsds,GHT93,BZ-str,BZ-p,IB07:M0}) are constrained spinors forming two rectangular blocks of the
$Spin(1,D-1)$ valued matrix corresponding to (the 'square root' of)
the $D$ dimensional moving frame matrix (\ref{harmUinAPP}). Their definition is $D$- and $p$- dependent,{\it i.e.} different
not only for different $D$ but also for different $p$-branes.
The harmonics appropriate to the description of massless
$D=11$ (super)particle are collected in  $Spin(1,10)$ valued matrix (\ref{harmVin}) obeying  Eqs.
(\ref{VGV=GU})--(\ref{VCV=C}). Its inverse matrix \cite{IB07:M0}
\begin{eqnarray}\label{harmVinAPP}
V_\alpha^{(\beta)}= (v_\alpha{}_q^{-}\; , v_\alpha{}_{q}^{+})\;  \in
\; Spin(1,10)
\end{eqnarray}
obeys
\begin{eqnarray}
\label{V-1:=APP} &  V_{(\beta)}{}^{\!\gamma} V_{\gamma}^{(\alpha)}=
 \delta_{(\beta)}^{\;(\alpha)}\quad \Leftrightarrow \quad
 \begin{cases} v^{+\alpha}_qv_{\alpha p}^{\; -}= \delta_{qp} & \; ,
 \qquad
v^{+\alpha}_qv_{\alpha p}^{\; +}= 0
 \cr v^{-\alpha}_qv_{\alpha p}^{\; -}= 0 & \; ,  \qquad    v^{-\alpha}_qv_{\alpha p}^{\; +}= \delta_{qp}
  \end{cases}
\end{eqnarray}
and
\begin{eqnarray}\label{harmV-1APP}
\begin{cases} V \Gamma^{(a)} V^T = \Gamma^{b} u_b^{(a)} \; , \quad
V^T \tilde{\Gamma}^{(a)} V =  u_b^{(a)} \tilde{\Gamma}^b\; ,  (a) \cr
V^TCV=C, \qquad VC^{-1}V^T=C^{-1} \; , \qquad (b)
\end{cases} \; . \qquad
\end{eqnarray}
The square root type relation between spinor moving
frame an moving frame variables encoded in the constraints (\ref{harmV-1APP}a) can be split further into
\begin{eqnarray}\label{vv=uG-all}
2v_\alpha{}^-_q v_\beta{}^-_q = \Gamma^{b}_{\alpha\beta} u_b^{=} \quad & (a) \; ,
\qquad
v^-_q\tilde{\Gamma}_bv^-_p = u_b^{=} \delta_{qp} \quad (d) & \; , \qquad \nonumber \\
2v_\alpha{}^+_q v_\beta{}^+_q = \Gamma^{b}_{\alpha\beta} u_b^{\#} \quad & (b) \; ,
\qquad
v^+_q\tilde{\Gamma}_bv^+_p = u_b^{\#} \delta_{qp} \quad (e) & \; , \qquad  \nonumber \\
2v_{(\alpha|}{}^+_q \gamma^i_{qp} v_{|\beta)}{}^+_p =
\Gamma^{b}_{\alpha\beta} u_b{}^{i} \quad & (c) \; , \qquad
v^-_q\tilde{\Gamma}_bv^+_p = u_b{}^{i } \gamma^i_{qp} \quad (f) & \;
. \qquad
\end{eqnarray}
The equations in (\ref{harmV-1APP}a), expressing
the Lorentz invariance of the  charge conjugation matrix $C$,
allow to construct (explicitly) the elements of the inverse spinor moving frame matrix, as in (\ref{V-1=CV}),
\begin{eqnarray}
\label{V-1=CV-APP}  v_{\alpha}{}^{\mp}_q = \pm i
C_{\alpha\beta}v_{q}^{\mp \beta }\; , \qquad  v^{\pm \alpha}_q = \pm i
C^{\alpha\beta}v_{\beta q}^{\; \pm}\; . \qquad
 \end{eqnarray}

In the massless superparticle
model the set of $16$ spinors $v_{\alpha p}^{\;-}$ in
(\ref{harmVinAPP}) can be identified with the homogeneous
coordinates of the celestial $S^9$ sphere given by the $SO(1,10)$
Lorentz group coset \cite{Ghsds,GHT93,IB07:M0} \begin{eqnarray}\label{v-inS11APP}
{} \{v_{\alpha p}^{\;-}\} = {Spin(1,10) \over [Spin (1,1)\otimes
Spin(9)] \, \subset \!\!\!\!\!\!\times \mathbb{K}_9 } =
\mathbb{S}^{9} \; .   \qquad
\end{eqnarray}
In the dynamical system of the massless (super)particle these describe the angles defining the direction of the light--like momentum so that one can consider $v_{\alpha p}^{\;-}$'s as carriers of all the momentum degrees of freedom but energy.
The set of others $16$ spinors, $v_{\alpha p}^{\;+}$, can be gauged away (but, of course, not set to zero) by the $K9$ transformations (\ref{M0:K9sym}), so that, in principle, one can work with the set of constrained spinors $v_{\alpha p}^{\;-}$ only. However, it is often convenient to use the complete spinor moving frame and keep only the  $SO(9)\otimes SO(1,1)$ symmetry as an equivalence relation on the set of $v_\alpha{}^-_q\,$'s and
$v_\alpha{}^+_q\,$'s which satisfy the set of constraints in (\ref{harmVinAPP}). Then these constrained spinorial variables become homogeneous coordinates of the non--compact $SO(1,10)/[SO(9)\times
SO(1,1)]$ coset, while $K9$ can be considered as a non-manifest (`hidden') symmetry.

The $SO(1,10)$ covariant derivatives $(d+w)$ of the harmonic variables which do not break the
kinematical constraints (\ref{harmUinAPP}), (\ref{harmVinAPP}) (admissible derivatives) are expressed by
\begin{eqnarray}
\label{du--=APP}  (d+w)u^{=}_a:= du^{=}_a+w_a{}^b u^{=}_b &=& - 2u^{=}_a \Omega^{(0)} + u^{i}_a
\Omega^{=i} \;  , \qquad
 \\
\label{du++}  (d+w)u^{\#}_a &=& +2u^{\#}_a \Omega^{(0)} + u^{i}_a
\Omega^{\#i} \; , \qquad \\
\label{dui}
 (d+w)u_a{}^{i}\; &=& {1\over 2}u^{=}_a \Omega^{\# i} + {1\over 2}u^{\#}_a \Omega^{= i}  -  u^{j}_a
\Omega^{ji} \; , \qquad
 \end{eqnarray}
through the covariant 1-forms
\begin{eqnarray}
\label{Omab=APP} \Omega^{(a)(b)}:= U^{c(a)}(d+w)U_c^{(b)} = - \Omega^{(b)(a)} =
\left(\begin{matrix}  0 & {1\over 2} ( \Omega^{\# i}+\Omega^{=i}) & 2\Omega^{(0)} \cr -
{1\over 2} ( \Omega^{\# i}+\Omega^{=i}) & \Omega^{ij} & - {1\over 2} ( \Omega^{\# i}-
\Omega^{=i}) \cr -2\Omega^{(0)} & {1\over 2} ( \Omega^{\# i}-\Omega^{= i}) & 0
\end{matrix}\right) \quad \;
\end{eqnarray}
which generalize the $SO(1,10)/[SO(1,1)\otimes SO(9)]$ Cartan forms for the case of local $SO(1,10)$ symmetry (see \cite{IB07:M0}).

As reflected by the constraints in (\ref{harmVinAPP}), the spinor
Lorentz harmonics $V$ (\ref{harmVinAPP}) give the spinor
representation of the Lorentz group element the vector
representation of which is given by the moving frame vectors
(\ref{harmUinAPP}). Then their admissible covariant derivatives ({\it i.e.}
derivatives preserving the constraints (\ref{harmVinAPP})) are
expressed through the same generalized Cartan forms
\begin{eqnarray}
\label{VdV=UdUG=APP} V_{(\alpha)}^{\gamma}(d+w)V_\gamma^{(\beta)} =   {1\over 4} \; \Omega^{(a)(b)}\;  \Gamma_{(a)(b)}{}_{(\alpha)}^{\;\; (\beta)}
\quad \in \; spin(1,10) \; , \qquad \Omega^{(a)(b)}:= U^{m(a)}(d+w)U_m^{(b)} \quad  \in \;
so(1,10)\; , \qquad
\end{eqnarray}
This implies
\begin{eqnarray}
\label{dv-q=APP}  (d+w)v_q^{-}=  - \Omega^{(0)} v_q^{-} - {1\over 4}
\Omega^{ij} v_p^{-}\gamma_{pq}^{ij} + {1\over 2} \Omega^{= i}
\gamma_{qp}^{i}v_p^{+} \; ,   \qquad
\\
 \label{dv+q} (d+w)v_q^{+} =  \Omega^{(0)} v_q^{+} -
{1\over 4} \Omega^{ij} v_p^{+}\gamma_{pq}^{ij} + {1\over 2} \Omega^{\# i}
\gamma_{qp}^{i}v_p^{-}  \;  \qquad
\end{eqnarray}
for the elements of the Spin(1,10) valued matrix (\ref{harmVinAPP}).

Since the Cartan forms $\Omega^{(0)}$ and $\Omega^{ij}$  transform
as connections  under local $SO(1,1)$ and $SO(9)$ transformations,
respectively, we can use them to define $SO(1,10)\otimes SO(1,1) \otimes SO(9)$ covariant exterior
derivatives (covariant differentials) of the moving frame variables. Unsing such a covariant
differential we can write   Eqs. (\ref{du--=APP}), (\ref{du++}),
(\ref{dv-q=APP}) and (\ref{dv+q}) in the form of
\begin{eqnarray}
\label{Du--=APP}  & Du^{=}_a :=(d+w)u^{=}_a + 2u^{=}_a \Omega^{(0)} =
u^{i}_a \Omega^{=i} \; , \qquad Du^{\#}_a :=(d+w)u^{\#}_a - 2u^{\#}_a
\Omega^{(0)} = u^{i}_a \Omega^{\#i} \; , \qquad \\ \label{Dui=APP} &
Du^{i}_a :=(d+w)u^{i}_a + u^{j}_m \Omega^{ji} = {1\over 2} u^{\#}_a
\Omega^{=i} + {1\over 2} u^{=}_a \Omega^{\#i} \; , \qquad
\end{eqnarray}
\begin{eqnarray}
\label{Dv-q=APP} &  Dv_q^{-} := (d+w)v_q^{-} + \Omega^{(0)} v_q^{-} +
{1\over 4} \Omega^{ij} v_p^{-}\gamma_{pq}^{ij} = {1\over 2}
\Omega^{=i} \gamma_{qp}^{i}v_p^{+} \; , \qquad
\\
 \label{Dv+q=APP} & Dv_q^{+} :=  (d+w)v_q^{+}  -
\Omega^{(0)} v_q^{+} + {1\over 4} \Omega^{ij}
v_p^{+}\gamma_{pq}^{ij} =  {1\over 2} \Omega^{\# i}
\gamma_{qp}^{i}v_p^{-} \; . \qquad
\end{eqnarray}

To simplify notation in (\ref{Dv-q=APP}) and (\ref{Dv+q=APP}) we have  omitted the spinorial indices; than it is not excessive to notice that in these equations  we have presented the covariant derivatives of the element of inverse spinor moving frame matrix (\ref{harmVinAPP}) carrying lower $Spin(1,10)$ index, while  the covariant derivatives of the initial spinor moving frame variables (\ref{harmVin}),  with upper $Spin(1,10)$ index, read
\begin{eqnarray}
\label{Dv-1-q=APP} &  Dv_q^{-\alpha} := dv_q^{-\alpha} +
\Omega^{(0)} v_q^{-\alpha} + {1\over 4} \Omega^{ij}
v_p^{-\alpha}\gamma_{pq}^{ij} = - {1\over 2} \Omega^{=i}
v_p^{+\alpha} \gamma_{pq}^{i}\; , \qquad \\
\label{Dv-1+q=APP} &  Dv_q^{+\alpha} := dv_q^{+\alpha} -
\Omega^{(0)} v_q^{+\alpha} + {1\over 4} \Omega^{ij}
v_p^{+\alpha}\gamma_{pq}^{ij} = - {1\over 2} \Omega^{\# i}
v_p^{-\alpha} \gamma_{pq}^{i}\; . \qquad
\end{eqnarray}

Also the following algebraic equations were useful in our calculations
\begin{eqnarray}
\label{v-G=} (v^-_q\Gamma_a)_\alpha = u^{=}_av_{\alpha}^{+q} - u^{i}_a\gamma^i_{qp}v_{\alpha}^{-p}
\; , \qquad (v^+_q\Gamma_a)_\alpha = u^{\#}_av_{\alpha}^{-q} - u^{i}_a\gamma^i_{qp}v_{\alpha}^{+p}
\; , \qquad \\
\label{v-G=} (\tilde{\Gamma}_a v^{-q})^\alpha = u^{=}_av_q^{+\alpha} + u^{i}_a\gamma^i_{qp}v_p^{-\alpha}
\; , \qquad  (\tilde{\Gamma}_a v^{+q})^\alpha = u^{\#}_av_q^{-\alpha} + u^{i}_a\gamma^i_{qp}v_p^{+\alpha}
\; , \qquad \\
\label{v-G=}  (v^-_q\Gamma_{ab}v^-_p)= 2 u^{=}_{[a}u^{i}_{b]}\gamma^i_{qp} \; . \qquad
\end{eqnarray}
\bigskip

\appendix{{\bf Appendix C: Some other technical details  }}
\renewcommand{\theequation}{C.\arabic{equation}}
\setcounter{equation}{0}
\bigskip

In calculating the $SO(9)$ curvature (\ref{M0:Gij=onshell}) from (\ref{RL=}) it is useful to notice that
$\widehat{*F}_{\# ijk_1...k_4} = {1\over 6}\epsilon^{= ijk_1...k_4l_1l_2l_3} \hat{F}_{\# l_1l_2l_3}$. Here and below the $SO(1,1)$ and $SO(9)$ indices are obtained by contraction with the moving frame variables $u^{=}_a$, $u^i_a$, $v^{-\alpha}_q$, Eqs.  (\ref{harmUin}) and (\ref{harmVin}).

Useful relations for the projections of the pull--back of the tensor--spin-tensor (\ref{ta:=}) to ${\cal W}^{(1|16)}$ are
\begin{eqnarray}\label{v-t--}
 (v^-_q\hat{t}^{=})^\alpha  = {1\over 36} \hat{F}_{\# ijk}\gamma^{ijk}_{qp}v_p^{+\alpha} \; , \qquad (\hat{t}^{=}v^-_q)_\alpha  = - {1\over 12} \hat{F}_{\# ijk}v_{\alpha}{}^+_p\gamma^{ijk}_{pq} \; , \qquad
 \\ \label{v-tiv-} (v^-_q\hat{t}^{i}v^-_p)  = - {1\over 6} \hat{F}_{\# jkl}\left(\delta^{i[j}\gamma^{kl]}_{qp}+ {1\over 6}
\gamma^{i jkl}_{qp}\right) \; . \qquad
\end{eqnarray}
The explicit form of the   tensor--spin-tensor in the last term of  Eq.  (\ref{DpT--iq=}) is
\begin{eqnarray} \label{Sigma:=}
 {\Sigma}{}^{i\, , \,  j_1j_2j_3\, ,\, k_1k_2k_3}_{pq} & := {1\over 18\cdot 4! } \left(
\gamma^{j_1j_2j_3} \left(\delta^{i[k_1}\gamma^{k_2k_3]}+ {1\over 6}
\gamma^{i k_1k_2k_3}\right) +  3 \left(\delta^{i[k_1}\gamma^{k_2k_3]}+ {1\over 6}
\gamma^{i k_1k_2k_3}\right)\gamma^{j_1j_2j_3} + \right. \qquad \nonumber \\
& \left. + (j_{1,2,3}\leftrightarrow k_{1,2,3})\right)\, . \qquad
\end{eqnarray}
It obeys
\begin{eqnarray} \label{SigmaProp}
 & {\Sigma}{}^{i\, , \,  j_1j_2j_3\, ,\, k_1k_2k_3}_{pp}=0\; , \qquad {\Sigma}{}^{i\, , \,  j_1j_2j_3\, ,\, k_1k_2k_3}_{qp}\gamma^{jk}_{pq}= 0\; , \qquad {\Sigma}{}^{i\, , \,  j_1j_2j_3\, ,\, k_1k_2k_3}_{qp}\gamma^i_{pq}= - {8\over 3} \delta^{[j_1}_{k_1}\delta^{j_2}_{k_2}\delta^{j_3]}_{k_3} \; , \qquad  \nonumber \\
& {\Sigma}{}^{i\, , \,  j_1j_2j_3\, ,\, k_1k_2k_3}_{qp}\gamma^j_{pq}= - {4\over 27} \left(9\delta^{i[j_1}\delta^{j_2}_{[k_2}\delta^{j_3]}_{k_3}\delta^{j}_{k_1]}+ \delta^{ij} \delta^{[j_1}_{k_1}\delta^{j_2}_{k_2}\delta^{j_3]}_{k_3} \right)\; . \qquad
\end{eqnarray}

\end{widetext}


\begin{thebibliography}{99}
\renewcommand{\theequation}{R.\arabic{equation}}


\bibitem{M-theory}
C.M.~Hull and P.~K.~Townsend,
Nucl.\ Phys.\ {\bf
B438}, 109-137 (1995) [hep-th/9410167]; E.~Witten,
Nucl.\ Phys.\  {\bf B443}, 85 -126 (1995) [hep-th/9503124];
J.H. Schwarz,
{Nucl.Phys.Proc.Suppl.} {\bf B55}, 1-32 (1997);
 P.K. Townsend, {\it
Four Lectures on M--theory}, hep-th/9612121.
In :{\sl Summer School High energy physics and cosmology,
Trieste 1996} (Eds. E. Gava, A. Masiero, K.S. Narain, S. Randjbar-Daemi, Q. Shafi), The
ICTP Series in Theoretical Physics, Vol. 13, World Scientific, 1997, Singapore,
pp. 385-438 [hep-th/9612121].


\bibitem{AdS/CFT}
J.M. Maldacena, {Adv. Theor. Math. Phys.} {\bf 2}, 231-252 (1998);
\,
S.S. Gubser, I.R. Klebanov and A.M. Polyakov,
{Phys.Lett.}
 {\bf B428}, 105-114 (1998);
\, E. Witten, {Adv. Theor. Math. Phys.} {\bf 2},
253-291 (1998);
\, O. Aharony, S.S. Gubser, J.
Maldacena, H. Ooguri and Y. Oz,
{Phys.Rept.} {\bf  323}, 183-386 (2000).

\bibitem{viscosity}
  G.~Policastro, D.~T.~Son and A.~O.~Starinets,
  ``The shear viscosity of strongly coupled N = 4 supersymmetric Yang-Mills
  plasma,''
  Phys.\ Rev.\ Lett.\  {\bf 87}, 081601 (2001)
  [arXiv:hep-th/0104066];
  P.~Kovtun, D.~T.~Son and A.~O.~Starinets,
  ``Viscosity in strongly interacting quantum field theories from black hole
  physics,''
  Phys.\ Rev.\ Lett.\  {\bf 94}, 111601 (2005)
  [arXiv:hep-th/0405231];
 D.~T.~Son and A.~O.~Starinets,
  ``Viscosity, Black Holes, and Quantum Field Theory,''
  Ann.\ Rev.\ Nucl.\ Part.\ Sci.\  {\bf 57}, 95 (2007)
  [arXiv:0704.0240 [hep-th]].


\bibitem{AppAdS/CFT}
 S.~A.~Hartnoll,
  ``Lectures on holographic methods for condensed matter physics,''
  Class.\ Quant.\ Grav.\  {\bf 26}, 224002 (2009)
  [arXiv:0903.3246 [hep-th]]; \\
   C.~P.~Herzog,
  J.\ Phys.\ A  {\bf 42}, 343001 (2009)
  [arXiv:0904.1975 [hep-th]]; \\
J.~P.~Gauntlett, J.~Sonner and T.~Wiseman,
  ``Holographic superconductivity in M-Theory,''
  Phys.\ Rev.\ Lett.\  {\bf 103}, 151601 (2009)
  [arXiv:0907.3796 [hep-th]];
J.~P.~Gauntlett, J.~Sonner and T.~Wiseman,
  ``Quantum Criticality and Holographic Superconductors in M-theory,''
  JHEP {\bf 1002}, 060 (2010)
  [arXiv:0912.0512 [hep-th]].

\bibitem{Howking}
Stephen Hawking, {\it G\"odel and the End of Physics}, public lecture on occasion of Dirac Centennial Celebration, Centre for Mathematical Sciences,
Wilberforce Road, Cambridge, July 20th 2002,   http://www.damtp.cam.ac.uk/strings02/dirac/hawking/.
Petr Ho\v{r}ava, privat communication 1997.



\bibitem{Tomas+}
  M.~Huebscher, P.~Meessen and T.~Ortin,
  ``Domain walls and instantons in N=1, d=4 supergravity,''
  arXiv:0912.3672 [hep-th].



\bibitem{Duff94-Stelle98}
  M.~J.~Duff, R.~R.~Khuri and J.~X.~Lu,
  ``String solitons,''
  Phys.\ Rept.\  {\bf 259}, 213 (1995)
  [hep-th/9412184];
  K.~S.~Stelle,
  ``BPS branes in supergravity,'' hep-th/9803116.
in:
{\it High Energy Physics and Cosmology  1997} (Eds: E. Gava et al.),
The ICTP Series in Theoretical Physics, Vol. 14,
Singapore, World Scientific, 1998,
pp. 29-127 [hep-th/9803116].

\bibitem{BST1987}
  E.~Bergshoeff, E.~Sezgin and P.~K.~Townsend,
  {\it  Supermembranes and eleven-dimensional supergravity},
  Phys.\ Lett.\  {\bf B189}, 75 (1987).



\bibitem{Dpac}
P.~K.~Townsend,
  {\it D-branes from M-branes},
  Phys.\ Lett.\  {\bf B373}, 68 (1996)
  [hep-th/9512062];
M.\ Cederwall, A.\ von Gussich, B.E.W.\ Nilsson, A.\ Westerberg,
{\it The Dirichlet super-three-branes in ten-dimensional type IIB
supergravity},
{Nucl.Phys.} {\bf B490}
(1997) 163--178  [hep-th/9610148];  \\
M.\ Aganagic, C.\ Popescu, J.H.\ Schwarz,
{\it D-brane actions with local kappa symmetry},
{Phys.Lett.} {\bf B393},  311--315 (1997)
[hep-th/9610249];
M. Cederwall, A.\ von Gussich, B.E.W.\ Nilsson, P.\ Sundell and A.\
Westerberg,
{\it The Dirichlet super-p-branes in ten-dimensional type IIA and IIB supergravity},
{Nucl.Phys.} {\bf B490}, 179--201 (1997)
[hep-th/9611159];
M.\ Aganagic, C.\ Popescu, J.H.\ Schwarz,
{\it Gauge--invariant and gauge--fixed D-brane actions},
{Nucl.Phys.} {\bf B490}, 202 (1997) [hep-th/9612080].

\bibitem{B+T=Dpac}
E.\ Bergshoeff, P.K.\ Townsend,
{\it Super-$D$-branes},
{Nucl.Phys.}
{\bf B490}, 145--162 (1997)  [hep-th/9611173].




\bibitem{blnpst}
  I.A. Bandos, K. Lechner, A. Nurmagambetov, P.~Pasti, D.P.~Sorokin, M.~Tonin,
 {\it Covariant action for the super-five-brane of M-theory},
  Phys.Rev.Lett. {\bf 78},  4332 (1997)
  [hep-th/9701149];
M.~Aganagic, J.~Park, C.~Popescu and J.H. Schwarz,
  {\it World-volume action of the M-theory five-brane},
  Nucl.\ Phys.\  {\bf B496}, 191-214 (1997)
  [hep-th/9701166].



\bibitem{bpstv}
  I. Bandos, D. Sorokin, M. Tonin, P. Pasti, D.V. Volkov,
  {\it Superstrings and supermembranes in the doubly supersymmetric geometrical
  approach},
  Nucl. Phys.  {\bf B446}, 79-118 (1995)
  [hep-th/9501113].



\bibitem{hs96}
 P.~S.~Howe and E.~Sezgin,
  {\it Superbranes},
  Phys.\ Lett.\  B {\bf 390}, 133 (1997).
  [hep-th/9607227].

\bibitem{hs2}
  P.~S.~Howe and E.~Sezgin,
  {\it D = 11, p = 5},
  Phys.\ Lett.  {\bf B394}, 62 (1997).
  [hep-th/9611008].


\bibitem{Dima99}
  D.~P.~Sorokin,
  {\it Superbranes and superembeddings},
  Phys.\ Rept.\  {\bf 329}, 1-101 (2000).


\bibitem{IB09:M-D}
  I.~A.~Bandos,
  ``Superembedding approach to Dp-branes, M-branes and multiple D(0)-brane
  systems,''
  arXiv:0912.2530 [hep-th].




\bibitem{stv}
  D.~P.~Sorokin, V.~I.~Tkach and D.~V.~Volkov,
  {\it Superparticles, twistors and Siegel symmetry},
  Mod.\ Phys.\ Lett.\  {\bf A4} (1989) 901-908;

\bibitem{stv+}
  N.~Berkovits, ``A Covariant Action For The Heterotic Superstring With Manifest Space-Time Supersymmetry And World Sheet Superconformal Invariance,´´
  Phys.\ Lett.\  {\bf B232}, 184 (1989); \\
  E.~A.~Ivanov and A.~A.~Kapustnikov,
  ``Towards a tensor calculus for kappa supersymmetry,''
  Phys.\ Lett.\  B {\bf 267}, 175 (1991); \\
 M.~Tonin,
  ``World sheet supersymmetric formulations of Green-Schwarz superstrings,''
  Phys.\ Lett.\  B {\bf 266}, 312 (1991); \\
  ``kappa symmetry as world sheet supersymmetry in D = 10 heterotic
  superstring,''
  Int.\ J.\ Mod.\ Phys.\  A {\bf 7}, 6013 (1992); \\
A.~Galperin and E.~Sokatchev,
  {\it A Twistor Like D = 10 Superparticle Action With Manifest N=8 Worldline
  Supersymmetry},
  Phys.\ Rev.\  {\bf D46}, 714-725 (1992)
  [hep-th/9203051]; \\
F.~Delduc, A.~Galperin, P.~S.~Howe and E.~Sokatchev,
  {\it A Twistor formulation of the heterotic D = 10 superstring with manifest
  (8,0) world sheet supersymmetry},
  Phys.\ Rev.\  {\bf 47}, 578-593 (1993)
  [hep-th/9207050]; \\
   I.~A.~Bandos, D.~P.~Sorokin, M.~Tonin and D.~V.~Volkov,
  Phys.\ Lett.\  B {\bf 319}, 445 (1993)
  [arXiv:hep-th/9307039]; \\
 D.~P.~Sorokin and M.~Tonin,
  ``On the Chiral fermions in the twistor - like formulation of D = 10 heterotic
  string,''
  Phys.\ Lett.\  B {\bf 326}, 84 (1994)
  [arXiv:hep-th/9307195]; \\
 E.~Ivanov and E.~Sokatchev,
  ``Chiral fermion action with (8,0) world sheet supersymmetry,''
  arXiv:hep-th/9406071; \\
 P.~S.~Howe,
  ``A Note On Chiral Fermions And Heterotic Strings,''
  Phys.\ Lett.\  {\bf B332}, 61 (1994)
  [arXiv:hep-th/9403177].


\bibitem{vz+}
 D.~V.~Volkov and A.~A.~Zheltukhin,
  ``On the equivalence of the Lagrangians of massless Dirac and supersymmetrical particles,''
  Lett.\ Math.\ Phys.\  {\bf 17}, 141 (1989); \\
D.~P.~Sorokin, V.~I.~Tkach, D.~V.~Volkov and A.~A.~Zheltukhin,
  ``From the Superparticle Siegel Symmetry to the Spinning Particle Proper Time
  Supersymmetry,''
  Phys.\ Lett.\ {\bf B216}, 302 (1989); \\
S.~Aoyama, P.~Pasti and M.~Tonin,
  ``The GS and NRS heterotic strings from twistor string models,''
  Phys.\ Lett.\  {\bf B283}, 213 (1992); \\
 D.~V.~Uvarov,
  ``On covariant kappa-symmetry fixing and the relation between the NSR  string
  and the type II GS superstring,''
  Phys.\ Lett.\  B {\bf 493}, 421 (2000)
  [arXiv:hep-th/0006185];
  Nucl.\ Phys.\ Proc.\ Suppl.\  {\bf 102}, 120 (2001)
  [arXiv:hep-th/0104235].




\bibitem{Witten:1995im}
  E.~Witten,
  {\it Bound states of strings and p-branes},
  Nucl.\ Phys.\   {\bf B460}, 335 (1996).


\bibitem{Tseytlin:DBInA}
  A.~A.~Tseytlin,
{\it On non-abelian generalisation of the Born-Infeld action in
string theory},
  Nucl.\ Phys.\ {\bf B501}, 41-52 (1997)
  [hep-th/9701125].

\bibitem{Dima01}
 D.~P.~Sorokin,
  {\it  Coincident (super)-Dp-branes of codimension one},
  JHEP {\bf 08}, 022 (2001)
  [hep-th/0106212];
  J.~M.~Drummond, P.~S.~Howe and U.~Lindstrom,
  {\it Kappa-symmetric non-Abelian Born-Infeld actions in three dimensions},
  Class.\ Quant.\ Grav.\  {\bf 19}, 6477 (2002)
  [hep-th/0206148];
S.~Panda and D.~Sorokin,
  {\it Supersymmetric and kappa-invariant coincident D0-branes},
  JHEP {\bf 0302},  055 (2003).






\bibitem{Howe+Linstrom+Linus}
 P.~S.~Howe, U.~Lindstrom and L.~Wulff,
 {\it Superstrings with boundary fermions},
  JHEP {\bf 0508}, 041 (2005)
  [hep-th/0505067];
\bibitem{Howe+Linstrom+Linus=2007}
P.~S.~Howe, U.~Lindstrom and L.~Wulff,
  {\it On the covariance of the Dirac-Born-Infeld-Myers action},
  JHEP {\bf 0702}, 070 (2007)
  [hep-th/0607156].





\bibitem{IB09:D0}
  I.~A.~Bandos,
  ``On superembedding approach to multiple D-brane system. D0 story,''
  Phys.\ Lett.\  {\bf B680}, 267--273 (2009)
  [arXiv:0907.4681 [hep-th]].


\bibitem{Myers:1999ps}
  R.~C.~Myers,
  {\it Dielectric-branes},
  JHEP {\bf 9912}, 022 (1999)
  [hep-th/9910053].

\bibitem{Marcus+Sagnotti=86}
  N.~Marcus and A.~Sagnotti,
  ``Group theory from quarks at the ends of strings,''
  Phys.\ Lett.\  B {\bf 188} (1987) 58.

\bibitem{mM0=PLB}
  I.A. Bandos,
 {\it Superembedding approach to M0-brane and multiple M0-brane system}, Phys. Lett. {\bf B687}, 258–-263 (2010) [arXiv:0912.5125[hep-th]].




\bibitem{Howe+Linstrom+Linus=2010}
  P.~S.~Howe, U.~Lindstrom and L.~Wulff,
  ``D=10 supersymmetric Yang-Mills theory at $\alpha^{\prime 4}$,''
  arXiv:1004.3466 [hep-th].





\bibitem{Schwarz:2004yj}
  J.~H.~Schwarz,
  ``Superconformal Chern-Simons theories,''
  JHEP {\bf 0411}, 078 (2004)
  [arXiv:hep-th/0411077].


\bibitem{Filippov} V.T. Filippov, ``n-Lie algebras'', Sib. Mat. Zh., {\bf 26},
No 6, 126-140 (1985).


\bibitem{BLG}
  J.~Bagger and N.~Lambert,
  ``Gauge Symmetry and Supersymmetry of Multiple M2-Branes,''
  Phys.\ Rev.\  {\bf D77} (2008) 065008
  [arXiv:0711.0955 [hep-th]];
 ``Comments On Multiple M2-branes,''
  JHEP {\bf 0802} (2008) 105
  [arXiv:0712.3738 [hep-th]];
  A.~Gustavsson,
 ``Algebraic structures on parallel M2-branes,''
  Nucl.\ Phys.\  B {\bf 811}, 66 (2009)
  [arXiv:0709.1260 [hep-th]].



\bibitem{ABJM}
  O.~Aharony, O.~Bergman, D.~L.~Jafferis and J.~Maldacena,
 ``N=6 superconformal Chern-Simons-matter theories, M2-branes and their
  gravity duals,''
  JHEP {\bf 0810}, 091 (2008)
  [arXiv:0806.1218 [hep-th]].


\bibitem{Iengo:2008cq}
  R.~Iengo and J.~G.~Russo,
  {\it Non-linear theory for multiple M2 branes},
  arXiv:0808.2473 [hep-th].


\bibitem{YLozano+=0207}
 B.~Janssen and Y.~Lozano,
 ``On the dielectric effect for gravitational waves,''
  Nucl.\ Phys.\  {\bf B643}, 399 (2002);
    ``A microscopical description of giant gravitons,''
  Nucl.\ Phys.\  {\bf B658}, 281 (2003)
  [hep-th/0207199].

\bibitem{Emparan:1997rt}
  R.~Emparan,
  {\it Born-Infeld strings tunneling to D-branes},
  Phys. Lett.  {\bf B423}, 71 (1998)
  [hep-th/9711106].


\bibitem{Banks:1996vh}
  T.~Banks, W.~Fischler, S.~H.~Shenker and L.~Susskind,
  {\it M theory as a matrix model: A conjecture},
  Phys.\ Rev.\  {\bf D55}, 5112-5128 (1997)
  [hep-th/9610043].

\bibitem{mM0=PRL}
Igor A.~Bandos,
  {\it   Multiple M-wave interaction with fluxes},
Phys. Rev. Lett. (2010) [to appear],  arXiv:1003.0399 [hep-th].


\bibitem{CremmerFerrara80}
E.~Cremmer and S.~Ferrara,
{\it Formulation of eleven-dimensional
supergravity in superspace},
Phys. Lett. {\bf B91}, 61 (1980).
\bibitem{BrinkHowe80}
L.~Brink and P.S.~Howe,
{\it Eleven-dimensional supergravity on the
mass-shell in superspace},
Phys. Lett. {\bf B91}, 384 (1980).

\bibitem{BdAPV05}
 I.~A.~Bandos, J.~A.~de Azc\'arraga, M.~Picon and O.~Varela, {\it On the formulation
of $D=11$ supergravity and the composite nature of its three-form field}, Ann. Phys.
{\bf 317} (2005) 238--279 [hep-th/0409100].

\bibitem{Susskind:1997cw}
  L.~Susskind,
  ``Another conjecture about M(atrix) theory,''
  arXiv:hep-th/9704080.



\bibitem{Seiberg:1997ad}
  N.~Seiberg,
  ``Why is the matrix model correct?,''
  Phys.\ Rev.\ Lett.\  {\bf 79}, 3577 (1997)
  [arXiv:hep-th/9710009].




\bibitem{Sok} E.~Sokatchev, {\it Light cone harmonic cuperspace cnd its
applications},
  Phys.\ Lett.\  {\bf B169}, 209-214 (1986);
  {\it Harmonic superparticle},
  Class.\ Quant.\ Grav.\  {\bf 4}, 237-246 (1987).

\bibitem{B90}
I.~A.~Bandos, {\it A Superparticle In Lorentz-Harmonic Superspace}, Sov.\ J.\ Nucl.\
Phys.\  {\bf 51}, 906--914 (1990) [Yad.\ Fiz.\ {\bf 51}, 1429--1444 (1990)];
{\it Multivalued action functionals, Lorentz
harmonics, and spin}, JETP Lett.\  {\bf 52}, 205 (1990).
 [Pisma Zh.\ Eksp.\
Teor.\ Fiz.\ {\bf 52}, 837 (1990)].

\bibitem{Ghsds}
  A.~S.~Galperin, P.~S.~Howe and K.~S.~Stelle, {\it The superparticle and the Lorentz
  group},  Nucl.\ Phys.\  {\bf B368}, 248-280 (1992)
  [hep-th/9201020];
\\ F.~Delduc, A.~Galperin and E.~Sokatchev,
  {\it Lorentz harmonic (super)fields and (super)particles},
  Nucl.\ Phys.\  B {\bf 368}, 143-171 (1992).




\bibitem{BZ-str}
I. A. Bandos and A. A. Zheltukhin, {\it Green-Schwarz
       superstrings in spinor moving frame formalism},
  Phys.\ Lett.\  {\bf B288}, 77-83 (1992);
{\it D = 10 superstring:
Lagrangian and Hamiltonian mechanics in twistor-like Lorentz harmonic formulation},
Phys.\ Part.\ Nucl.\ {\bf 25} (1994) 453-477 [Preprint IC-92-422, ICTP, Trieste, 1992,
81pp.]

\bibitem{GHT93}
 A.~S.~Galperin, P.~S.~Howe and P.~K.~Townsend,
  {\it Twistor transform for superfields},
  Nucl.\ Phys.\  {\bf B402}, 531 (1993).



\bibitem{BZ-p}
 I.~A.~Bandos and A.~A.~Zheltukhin,
  {\it Generalization of Newman-Penrose dyads in connection with the action
  integral for supermembranes in an eleven-dimensional space},
  JETP Lett.\  {\bf 55}, 81 (1992);
  {\it Eleven-dimensional supermembrane in a spinor moving repere
  formalism},
  Int.\ J.\ Mod.\ Phys.\   {\bf
   A8}, 1081--1092 (1993);
  {\it N=1 superp-branes in twistor--like Lorentz harmonic
  formulation},
  Class.\ Quant.\ Grav.\  {\bf 12}, 609-626  (1995)
  [hep-th/9405113].

\bibitem{IB07:M0}
  I.~A.~Bandos,
  ``Spinor moving frame, M0-brane covariant BRST quantization and intrinsic
  complexity of the pure spinor approach,''
  Phys.\ Lett.\  B {\bf 659}, 388 (2008)
  [arXiv:0707.2336 [hep-th]];
  ``D=11 massless superparticle covariant quantization, pure spinor BRST charge
  and hidden symmetries,''
  Nucl.\ Phys.\  B {\bf 796}, 360 (2008)
  [arXiv:0710.4342 [hep-th]].


\bibitem{Cederwall:2001td}
  M.~Cederwall, B.~E.~W.~Nilsson and D.~Tsimpis,
  ``D = 10 super-Yang-Mills at O(alpha**2),''
  JHEP {\bf 0107}, 042 (2001)
  [arXiv:hep-th/0104236].


\bibitem{Movshev:2009ba}
  M.~Movshev and A.~Schwarz,
  ``Supersymmetric Deformations of Maximally Supersymmetric Gauge Theories.
  I,''
  arXiv:0910.0620 [hep-th].


\bibitem{Hoppe82+}
  B.~de Wit, J.~Hoppe and H.~Nicolai, ``On the quantum mechanics of supermembranes,''
  Nucl.\ Phys.\  B {\bf 305}, 545 (1988);



\bibitem{NahmEq}
  W.~Nahm,
  ``A Simple Formalism For The Bps Monopole,''
  Phys.\ Lett.\  B {\bf 90}, 413 (1980).


\bibitem{BasuHarvey}
  A.~Basu and J.~A.~Harvey,
  ``The M2-M5 brane system and a generalized Nahm's equation,''
  Nucl.\ Phys.\  B {\bf 713}, 136 (2005)
  [arXiv:hep-th/0412310].





\bibitem{Dima+Linus+09}
  J.~Gomis, D.~Sorokin and L.~Wulff,
  ``The complete AdS(4) x CP(3) superspace for the type IIA superstring and
  D-branes,''
  JHEP {\bf 0903}, 015 (2009)
  [arXiv:0811.1566 [hep-th]];
  \\
  P.~A.~Grassi, D.~Sorokin and L.~Wulff,
  ``Simplifying superstring and D-brane actions in AdS(4) x CP(3)
  superbackground,''
  JHEP {\bf 0908}, 060 (2009)
  [arXiv:0903.5407 [hep-th]].

\bibitem{BMN}
  D.~E.~Berenstein, J.~M.~Maldacena and H.~S.~Nastase,
  ``Strings in flat space and pp waves from N = 4 super Yang Mills,''
  JHEP {\bf 0204}, 013 (2002)
  [arXiv:hep-th/0202021].

\bibitem{SheikhJabbari:2004ik}
  M.~M.~Sheikh-Jabbari,
  ``Tiny graviton matrix theory: DLCQ of IIB plane-wave string theory, a
  conjecture,''
  JHEP {\bf 0409}, 017 (2004)
  [arXiv:hep-th/0406214];
 M.~Torabian,
  ``Matrix Theory for the DLCQ of Type IIB String Theory on the
  AdS/Plane-wave,''
  Phys.\ Rev.\  D {\bf 76}, 026006 (2007)
  [arXiv:hep-th/0701046].

\bibitem{Feinstein:2002ay}
  A.~Feinstein,
  ``Penrose limits, the colliding plane wave problem and the classical  string
backgrounds,''
  Class.\ Quant.\ Grav.\  {\bf 19}, 5353 (2002)
  [arXiv:hep-th/0206052].

\bibitem{MatrixString}
  R.~Dijkgraaf, E.~P.~Verlinde and H.~L.~Verlinde,
  ``Matrix string theory,''
  Nucl.\ Phys.\  B {\bf 500}, 43 (1997)
  [arXiv:hep-th/9703030]; \\
  Y.~Kitazawa and S.~Nagaoka,
  ``Green-Schwarz superstring from type IIB matrix model,''
  Phys.\ Rev.\  D {\bf 77}, 026009 (2008)
  [arXiv:0708.1077 [hep-th]]
and refs. therin.

\bibitem{Nastase-ABJM}
  A.~Mohammed, J.~Murugan and H.~Nastase,
  ``Looking for a Matrix model of ABJM,''
  arXiv:1003.2599 [hep-th].




\end{thebibliography}
\end{document}